\documentclass[aps,prb,twocolumn,floatfix,amsmath,amssymb,superscriptaddress]{revtex4-1}
\pdfoutput=1
\usepackage{times}
\usepackage{bbold}
\usepackage{graphicx}
\usepackage{dcolumn}
\usepackage{bm}
\usepackage{color}
\usepackage{hyperref}

\begin{document}

\title{ Majorana zero modes  in a quantum Ising chain with longer-ranged interactions}
\author{Yuezhen Niu}
\altaffiliation[Currently at ]{School of Physics, Peking University, Beijing 100871, China}
\affiliation{Department of Physics and Astronomy, University of
California Los Angeles, Los Angeles, California 90095-1547, USA}
\author{Suk Bum Chung}
\affiliation{Stanford Institute for Theoretical Physics, Stanford, California 94305, USA}
\author{Chen-Hsuan Hsu}
\affiliation{Department of Physics and Astronomy, University of
California Los Angeles, Los Angeles, California 90095-1547, USA}
\author{Ipsita Mandal}
\affiliation{Department of Physics and Astronomy, University of
California Los Angeles, Los Angeles, California 90095-1547, USA}
\author{S. Raghu}
\affiliation{Department of Physics, Stanford University, Stanford, California 94305, USA}
\author{Sudip Chakravarty}
\affiliation{Department of Physics and Astronomy, University of
California Los Angeles, Los Angeles, California 90095-1547, USA}

\date{\today}

\begin{abstract}
A one-dimensional Ising model in a transverse field can be mapped onto a system of spinless fermions with  $p$-wave superconductivity.  In the weak-coupling BCS regime, it
exhibits a zero energy  Majorana mode at each end of the chain. Here, we consider a variation of the model, which 
represents a superconductor with 
longer ranged kinetic energy and pairing amplitudes, as is likely to occur in more realistic systems.
It possesses a richer 
zero temperature 
phase diagram and has several quantum phase transitions.
From an exact solution of the model these phases can be classified according to the number of Majorana zero modes of an open chain: 0, 1, or 2 at each end.  The model posseses a multicritical point where phases with 0, 1, and 2 Majorana end modes meet.  The number of Majorana modes at each end of the chain is identical to the topological winding number of the Anderson's pseudospin vector that describes the BCS Hamiltonian.  The topological classification of the phases requires a unitary time-reversal symmetry to be present.  When this symmetry is broken, only the number of Majorana end modes modulo 2 can be used to distinguish two phases. 
In one of the regimes,  the wave functions of the  two phase shifted Majorana zero modes decays exponentially in space  but in an oscillatory manner. The wavelength of oscillation is identical  to the  asymptotic connected  spin-spin correlation of the $XY$-model in a transverse field to which our model is dual.
\end{abstract}

\maketitle

\section{Introduction}
There has been much recent interest in Majorana zero modes.~\cite{wilczek:2009,Moore:1991,Nayak:1996,Ivanov:2001,Fu:2008,Sau:2010,Sau:2010b,Alicea:2010,Lutchyn:2010,Oreg:2010,Roy:2010,Bena:2011,DeGottardi:2011} Their   relevance to topologically protected
 quantum computation is intensely studied.   Kitaev~\cite{Kitaev:2001} suggested an elegant model of a one-dimensional $p$-wave superconducting wire, which supports Majorana zero modes at the ends of the chain. 
 
 Kitaev's model  is the fermionized version of the familiar one-dimensional   transverse field Ising model (TFIM),~\cite{Pfeuty:1970} which is   one of the simplest models of  quantum criticality. 
 In the fermionic representation, the well-known quantum phase transition in the model can be understood as a transition from the weak pairing BCS regime to the strong pairing BEC regime.~\cite{Read:2000}  The weak-pairing phase is topologically non-trivial and in this phase the chain with open boundaries possesses a Majorana fermion zero enegy mode localized at each end.  It is equivalent to the ferromagnetic phase of the transverse field Ising chain.  The strong-pairing phase is topologically trivial and does not have any normalizable Majorana fermion zero enegy modes at the ends.  It corresponds to the quantum disordered phase of the transverse field Ising chain.  Recently, there have been attempts to realize Kitaev's model in one dimensional wire networks.~\cite{Alicea:2011}  In a realistic quantum wire, however, the range of the hybridization of the electron wave function, as well as that of Cooper pairing will be of finite range and the effect of such longer ranged interactions must be addressed.  The goal of this paper is to study the effect of such longer ranged interactions.  We do so within the context of another exactly solvable model, and find a rich phase diagram that results from such longer ranged correlations. 
 
These longer ranged interactions were considered in a previously introduced generalization of the Ising model in a transverse field by extending it to contain   a  three spin interaction term,~\cite{Kopp:2005} which is also exactly solved by a Jordan-Wigner transformation.~\cite{Pfeuty:1970} This generalized model  
arises as a first step of a real space renormalization group transformation~\cite{Hirsch:1979} and has a  richer phase diagram. The purpose of that study was to understand how irrelevant operators can drive a system along a critical line between two different zero temperature quantum critical points. The flow of this crossover as the higher energy states are integrated out conform to the Zamolodchikov's $c$-theorem.~\cite{Neto:1993} It is an explicit example of an exactly solved case where a system that points to a given fixed point at higher energies can asymptotically flow to a different fixed point at lower energy scales. This flow was explicitly traced in terms of a flow from higher temperature to lower temperature. The lesson learnt there was that at higher temperatures a system may be pointing to a different fixed point  compared to its true fate at zero temperature.

Here we reexamine the Ising model with a transverse field, with the added three spin interaction from the perspective of  Majorana zero modes. We find that the phase diagram can be classified according to the number of Majorana zero modes. The fermionized version of this model corresponds to a p-wave superconductor in which the electrons have longer ranged hoppings and longer ranged harmonics of the p-wave gap function, enabling us to address the effect of such longer ranged interactions on the zero temperature phase diagram of the quantum Ising chain.  We find several topological phase transitions in our model, and the phases can be classified by a topological invariant of the Anderson pseudospin vector~\cite{Anderson:1958} of the mean-field description of the superconducting state.  This topological invariant is an integer, $\mathbb{Z}$, and also specifies the number of normalizable Majorana fermion zero energy modes that are localized at each end of a chain with open boundary conditions. 

 We assess the conditions under which the topological order of the zero temperature phase diagram remains intact, and find that all phases are protected by a unitary version of time-reversal symmetry (appropriate for spinless fermions): so long as this time reversal symmetry is preserved, the phases described in our work remain stable.  In particular, we find that even when there are 2 Majorana fermion zero modes localized at {\it each} end, separated by a lattice spacing with wave functions orthogonal to each other.  Once we allow breaking of  time reversal invariance the  topological invariant collapses to $\mathbb{Z}_{2}$, which implies at most one Majorana zero mode at each end of  an open chain. The results of the 1D superconductor also offers insights into TFIM. We find that the phase dominated by the three spin interaction has ground state degeneracy from analyzing the Majorana zero modes. However, we also find that there exists a class of local spin interaction that can remove this ground state degeneracy. Such impurities are time reversal breaking and it is not clear how they  realize in generic circumstances. 

The dual~\cite{Fradkin:1978} (exchanging  the site spins  by the bond spins)  of the three-spin model that we study is amusingly  the one-dimensional quantum $XY$-model in a transverse magnetic field~\cite{Barouch:1971} in an enlarged parameter space than studied previously. From a complex calculation of the asymptotic form of the  connected $z$-component of the   instantaneous spin-spin correlation function it was discovered that there is an oscillatory region within the ferromagnetic phase. We find that this phenomenon of oscillation is intimately related to the oscillation of the Majorana zero modes, most remarkably the oscillation wavelengths are identical, as is the exponential decay in the vicinity of the quantum critical lines. 

The plan of the paper is as follows: in Sec. II we set the stage by recapitulating the phase diagram of the model to orient the reader. In Sec. III Majorana zero modes and their properties are obtained from the solution of the Bogoliubov-de Gennes (BdG)  equation, while in Sec. IV we discuss the efficacy of the Majorana representation by obtaining the solution of a three-term recursion relation instead of the full numerical solution of the BdG Hamiltonian. In Sec. V we discuss the topological aspects and Sec. VI is the concluding section. There are three appendices giving some details.

\section{The Hamiltonian and the phase diagram}
The three spin extension of the TFIM, which was previously studied,~\cite{Kopp:2005} has the Hamiltonian
\begin{equation}
H=-\sum_i(g \sigma_i^x+\lambda_2\sigma_i^{x}\sigma_{i-1}^z\sigma_{i+1}^z+\lambda_1\sigma_i^z\sigma_{i-1}^z)
\end{equation}
The $\sigma$'s are the standard Pauli matrices.  
In this section we introduce the Hamiltonian and its phase diagram from a conventional Jordan-Wigner analysis.  The Hamiltonian 
 after Jordan-Wigner
transformation 
\begin{eqnarray}
\sigma_i^x = 1-2c_i^\dagger c_i \\
\sigma_i^z=-\prod_{j<i}(1-2c_j^\dagger c_j)(c_i +c_i^\dagger)
\end{eqnarray}
is
\begin{equation}
\label{eq:Ham1}
\begin{split}
H&=-g\sum_{i=1}^{N}(1-2c_i^\dagger
c_i)-\lambda_1\sum_{i=1}^{N-1}(c_i^\dagger c_{i+1}+c_i^\dagger
c_{i+1}^\dagger+h.c.)\\&-\lambda_2\sum_{i=2}^{N-1}(c_{i-1}^\dagger
c_{i+1}+c_{i+1}c_{i-1}+h.c.).
\end{split}
\end{equation}
In contrast to the spin model, the spinless fermion Hamiltonian is actually a one-dimensional {\em mean field} model for a triplet superconductor, where there are both nearest and next nearest neighbor hopping,
as well as condensates. The nearest neighbor hopping amplitude $\lambda_{1}$ is also the amplitude of the nearest neighbor superconducting gap, and the 
next nearest neighbor hopping amplitude is equal to the next nearest neighbor superconducting gap; in general $\lambda_{1}\ne \lambda_{2}$. 
In terms  of Jordan-Wigner fermions one can envision finding an actual one-dimensional system with such an extended Hamiltonian. The solution of the corresponding spin Hamiltonian through Jordan-Wigner transformation  is, however, exact and includes all possible fluctuation effects and is {\it not a mean field solution}.

\begin{figure}[htbp]
\begin{center}
\includegraphics[scale=0.35]{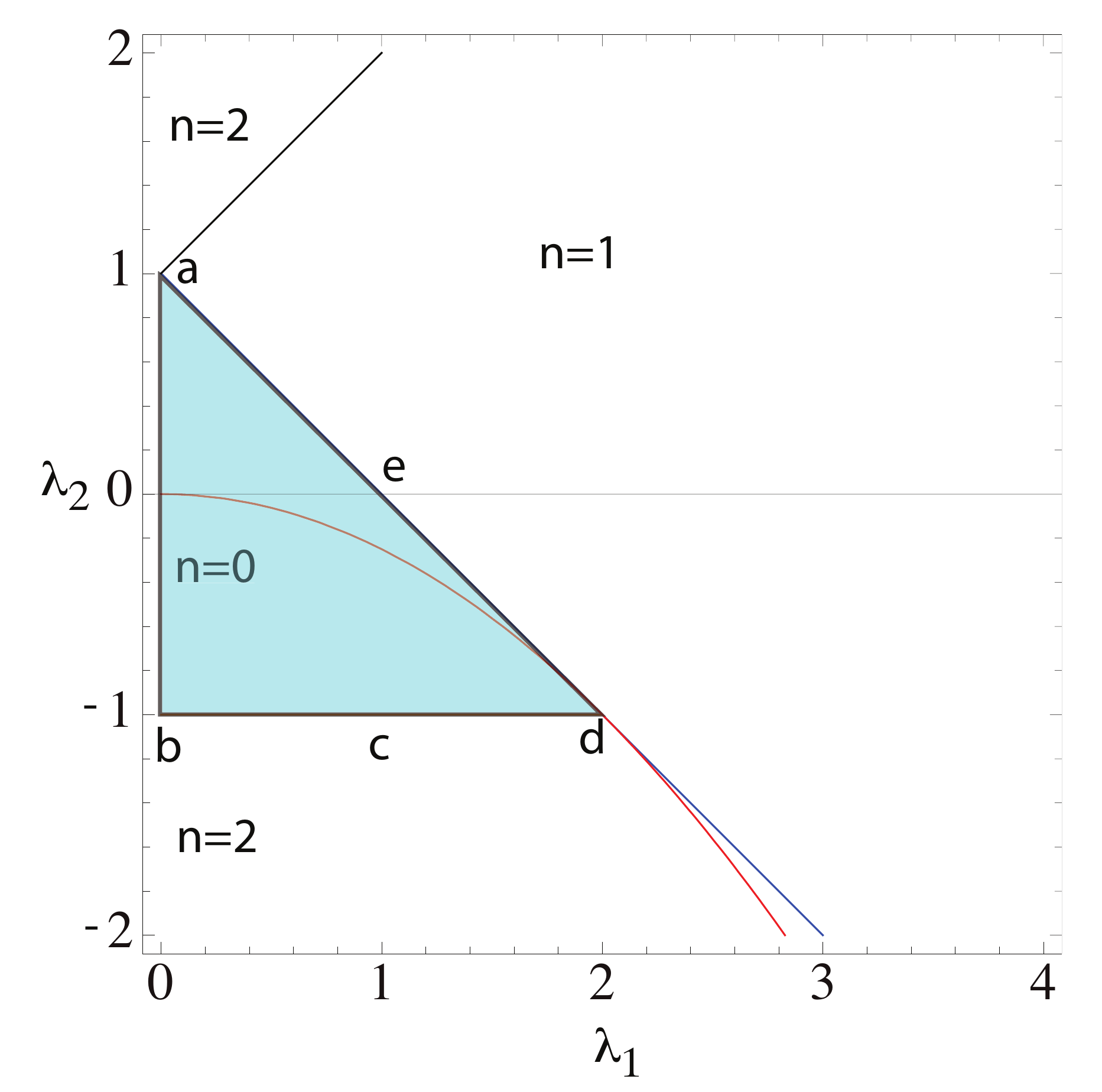}
\caption{(Color online)The region $\lambda_{2}>1+\lambda_{1}$ is ordered in the original spin representation and the boundary of it is a critical line where the gap at $k=0$ collapses. The region $\lambda_{2}<1-\lambda_{1}$ is disordered as well and the boundary corresponds a critical line where the gap at $k=\pi$ collapses. The point $\lambda_{1}=0$ and $\lambda_{2}=1$ is a special multicrtical point with an emergent $U(1)$ symmetry, most transparently seen in the dual representation (see below). In the same dual representation, the region enclosed by $\lambda_{1}^{2}=-4\lambda_{2}$ is an oscillatory ferromagnetically ordered phase separating
from an ordered phase for $\lambda_{2}<0$, as determined by the spatial decay of the instantaneous spin-spin correlation function. Note that duality exchanges ordered and disordered phases. Here $n=0, 1, 2$ correspond to  regions with $n$-Majorana zero modes at each end of an open chain.}
\label{fig:fig1}
\end{center}
\end{figure}

Imposing periodic boundary condition, the Hamiltonian can be immediately diagonalized by a Bogoliubov transformation:
\begin{equation}
H=\sum_{k} \varepsilon_{k}\left(\eta_{k}^{\dagger}\eta_{k}-\frac{1}{2}\right).
\end{equation}
The anticommuting fermion operators $\eta_{k}$'s are suitable linear combinations in the momentum space of the original Jordan-Wigner fermions. The spectra of excitations are (lattice spacing will be set to unity  throughout the paper)
\begin{equation}
\varepsilon_{k}=\pm 2\sqrt{1+\lambda_{1}^{2}+\lambda_{2}^{2}+2\lambda_{1}(1-\lambda_{2})\cos k -2\lambda_{2}\cos 2k}
\end{equation}
unless, otherwise stated, we shall set $g=1$.
Quantum phase transitions of this model are given by the nonanalyticities of the ground state energy:
\begin{equation}
E_{0}=-\frac{1}{2}\sum_{k}\varepsilon_{k}.
\end{equation}
These nonanalyticites are also defined by the critical lines where the gaps collapse; see Fig.~\ref{fig:fig1}. 

For the Ising model in a transverse field without  three spin interaction,
the gaps collapse at the Brillouin zone boundaries, $k=\pm \pi$ at the self-dual point $\lambda_{1}=1$ and $\lambda_{2}=0$. When the three spin interaction is added, the gaps can 
collapse at $k=0$ as well as $k=\arccos(\lambda_{1}/2)$ for  $\lambda_{2}=-1$ and $0<\lambda_{1}<2$.
 But at the free fermion point $b$,  there are no zero energy excitations except at $k=\pm \pi/2$. When we move to the point $c$ the spectrum evolves increasing  the weight at $k=0$ and the locations of the nodes are incommensurate with the lattice. The incommensuration shifts as a function of $\lambda_{2}$. The point $d$ 
 is a multicritical point and the spectra vanishes quadratically at $\pm \pi$. As we shall discuss below, the spectra are no longer relativistic at this point as a result of the confluence of two Dirac points,  corresponding to a dynamical exponent $z=2$.

\section{Majorana zero modes}
In this section we explore the zero modes by the Bogoliubov-de Gennes equations with open boundary condition.
\subsection{Unbroken  time reversal invariance}
\label{sec.secIIIA}
 The  equations, assuming open boundary condition, are given by
\begin{equation}
\left( \begin{array}{cc}
\hat{h}&\hat{\Delta}\\
-\hat{\Delta}&-\hat{h}\\
\end{array} \right)
\left(\begin{array}{c} \vec{u}_{n}\\\vec{v}_{n}\end{array}\right) = 
E_n\left(\begin{array}{c} \vec{u}_{n}\\\vec{v}_{n}\end{array}\right),
\end{equation}
where the submatrices are (unless otherwise stated, we will set $g=1$)
\begin{eqnarray}
\hat{h}_{ij}&=&\lambda_1(\delta_{j,i+1}+\delta_{j,i-1})+\lambda_2(\delta_{j,i+2}+\delta_{j,i-2})-2\delta_{ij}\\
\hat{\Delta}_{ij}&=&-\lambda_1(\delta_{j,i+1}-\delta_{j,i-1})-\lambda_2(\delta_{j,i+2}-\delta_{j,i-2})
\end{eqnarray}
Here $\vec{u}_{n}^{T}=(u_{n}(1), u_{n}(2),\dots u(N))$ and from time reversal symmetry  of the Hamiltonian $\vec{u}_{n}=\vec{u}_{n}^{*}$ and $\vec{v}_{n}=\vec{v}_{n}^{*}$. The  eigenvalue is labeled by $n$ and the arguments of $\vec{u}$ and  $\vec{v}$ are lattice indices.

From the diagonaliztion of the BdG Hamiltonian, we can easily see that the Majorana zero-modes can occur only for open boundary condition.
The phase diagram  itself can be deduced from the number of  zero modes of the BdG equation. In the next section we shall see that in the Majorana representation, the zero modes can be obtained from  a very simple recursion relation. 
With reference to Fig.~\ref{fig:fig1}
we note that there are regions of $n=0$, $n=1$, and $n=2$ zero modes, and the lines  separating them are quantum critical lines, except for the line separating $n=0$ and $n=2$, which is a topological transition.
The thin line $\lambda_{1}^{2}=-4\lambda_{2}$  corresponds to zero entanglement entropy.~\cite{Wei:2011} In this respect, it is remarkable that this thin line osculates the quantum critical line.  

\begin{figure}[h]
\includegraphics[width=\linewidth]{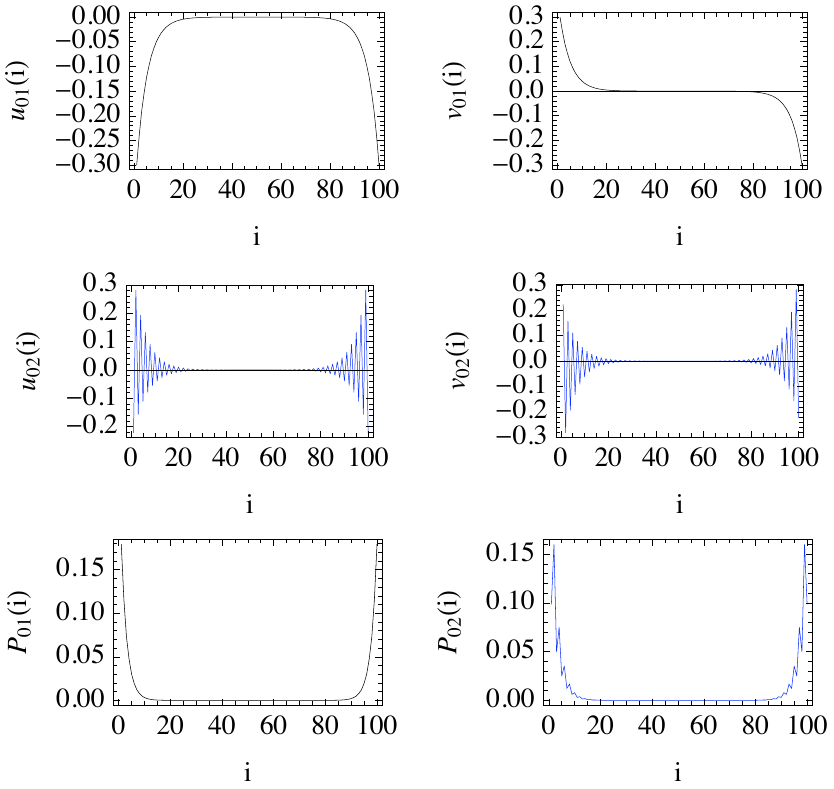}
\caption{(Color online)The two Majorana zero modes for $\lambda_{1}=0.05$ and $\lambda_{2}=1.5$. The row number 1 shows $u_{01}(i)$ and $v_{01}(i)$ corresponding to the the first Majorana zero mode. The second row corresponds to the second Majorana zero mode that is orthogonal to the first. The third row corresponds to the probability distributions $P_{01}(i)$ and $P_{02}(i)$ for the two respective Majorana modes. The numerical diagonalization was carried out for a lattice of $N=100$ sites with open boundary condition. For lattices larger than $150$, one quickly looses numerical control because of  $10^{16}$ difference in the order of magnitudes of the largest eigenvalue and the zero mode.}
\label{fig:fig5}
\end{figure}

For $\lambda_{2}>0$, and $\lambda_{2} >1 + \lambda_{1}$, one of the zero modes  decays exponentially in the bulk, and the decay length diverges as the quantum critical line is approached. The amplitude of the second zero mode also decays exponentially but it oscillates as $e^{i\pi n}$ regardless of $\lambda_{1}$ as it approaches the quantum critical line, at which point it loses the compactness of its support, signifying the loss of this zero mode. On the side $\lambda_{2}< 1 + \lambda_{1}$, one zero mode is recovered and it decays exponentially in the bulk, as in the region $\lambda_{2}>1 + \lambda_{1}$, as shown in Fig.~\ref{fig:fig5}. 

\begin{figure}[h]
\includegraphics[width=\linewidth]{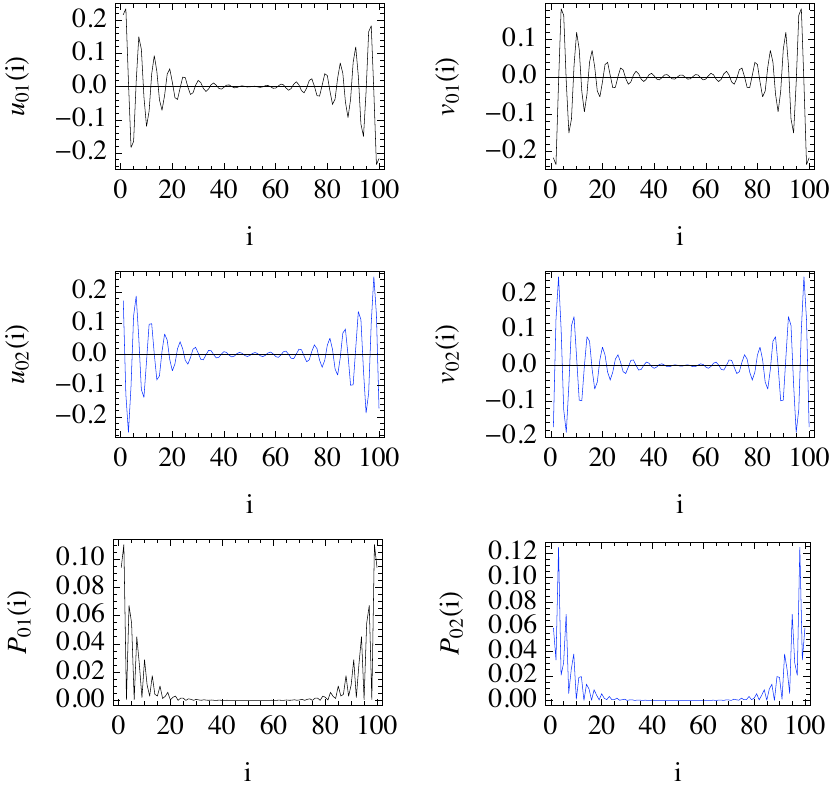}
\caption{(Color online)The two Majorana zero modes for s $\lambda_{2}= -1.2 $ and $\lambda_{1}=1$. The row number 1 depicts  $u_{02}(i)$ and $v_{02}(i)$ corresponding to the the first Majorana zero mode. The second row shows the second Majorana zero mode that is orthogonal to the first. The third row corresponds to the probability distributions $P_{01}(i)$ and $P_{02}(i)$ for the two respective Majorana modes.}
\label{fig:fig6}
\end{figure}

The situation is richer for $\lambda_{2}<0$. First of all there are no zero  modes until $\lambda_{2}<-1$ for $\lambda_{2}<1-\lambda_{1}$. In this region, there are two zero modes, both of which are oscillatory with exponentially decaying envelope. But this time the wavelengths of  the modes depend on the parameters $(\lambda_{1},\lambda_{2})$. Note that they are phase shifted by a lattice site; see Fig.~\ref{fig:fig6}. When we cross the quantum critical line, $\lambda_{2}=1-\lambda_{1}$, a non-oscillatory and exponentially decaying zero mode is observed.
\subsection{Broken time reversal invariance}
\label{sec:secIIIB}
We can ask what happens if we add a relative phase between the two order parameters  in the BdG Hamiltonian, while keeping the single particle Hamiltonian intact. Then,
\begin{equation}
\left( \begin{array}{cc}
\hat{h}&\hat{\Delta}\\
\hat{\Delta}^{\dagger}&-\hat{h}\\
\end{array} \right)
\left(\begin{array}{c} \vec{u}_{n}\\\vec{v}_{n}^{*}\end{array}\right) = 
E_n\left(\begin{array}{c} \vec{u}_{n}\\\vec{v}_{n}^{*}\end{array}\right),
\end{equation}
where the submatrices are 
\begin{eqnarray}
\hat{h}_{ij}&=&\lambda_1(\delta_{j,i+1}+\delta_{j,i-1})+\lambda_2(\delta_{j,i+2}+\delta_{j,i-2})\nonumber\\
&-&2\delta_{ij}\\
\hat{\Delta}_{ij}&=&e^{i \theta } \lambda_{2} (\delta _{i,j+2}-\delta _{i+2,j})+\lambda_{1} (\delta _{i-1,j}-\delta
   _{i+1,j})
   \label{eq:complexBdG}
\end{eqnarray}
The solution of this modified BdG Hamiltonian shows that the regions of the phase diagram which contain 
$n=1$  Majorana zero mode remain robust while those containing $n=2$ Majorana zero modes are in general 
destroyed, meaning that they are split; two examples are shown below in Fig.~(\ref{fig:split-Majorana-1})  for $\theta=\pi/2$.
For an arbitrary value of $\theta$, the real and the imaginary  parts of $\hat{\Delta}_{ij}$ will receive contributions from both 
$\lambda_{1}$ and $\lambda_{2}$ making it difficult to directly compare   with the 
previous phase diagram in Fig.~\ref{fig:fig1}. For $\theta=\pi/2$, the term containing $\lambda_{2}$ will be purely imaginary, while the $\lambda_{1}$ term
will remain untouched. This is easier to compare with the previous phase diagram because the absolute magnitude of the next-nearest
neighbor condensate remains the same. We have verified that our conclusions hold for arbitrary $\theta$ as well.
\begin{figure}[h]
\includegraphics[width=\linewidth]{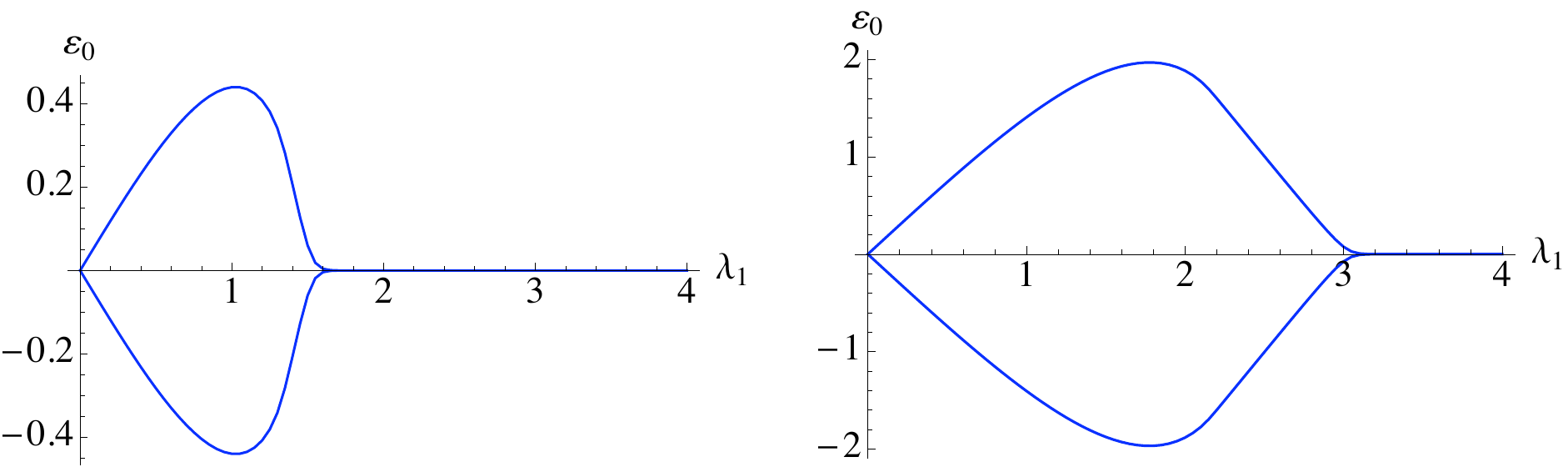}
\caption{(Color online)Splitting of $n=2$ Majorana zero modes for the complex Hermitian BdG equation. Left:  $\lambda_{2}=2.5$. Right: $\lambda_{2}=-2.0$.  The magnitude of $\varepsilon_{0}$ is the smallest eigenvalue. The slight rounding in the proximity of the quantum critical point is due to the finite size of the lattice: $N=200$. The quantum critical point for the infinite system is at $\lambda_{1}=1.5$ for the left panel and $\lambda_{1}=3.0$ for the right panel. The $n=1$ Majorana zero mode survives intact. Note that at $\lambda_{1}=0$ the chain splits into two independent chains and hence there is a zero mode irrespective of the fact that $\lambda_{2}$ pairing amplitude is purely imaginary.}
\label{fig:split-Majorana-1}
\end{figure}

\section{Majorana Hamiltonian}
To unveil the
hidden topology behind the TFIM with three spin
interaction, we introduce Majorana fermion operators:
\begin{eqnarray}
a_i = c_i^\dagger+c_i, \\
b_i=-i( c_i^\dagger-c_i), 
\end{eqnarray}
satisfying anti-commutation relations, so that the Hamiltonian in Eq.~(\ref{eq:Ham1}) becomes
\begin{equation}
H=-i\left[-\sum_{i=1}^Nb_ia_i+\lambda_1\sum_{i=1}^{N-1}b_i
a_{i+1}+\lambda_2\sum_{i=2}^{N-1}b_{i-1} a_{i+1}\right].
\end{equation}
The  three-spin interaction corresponds
to a next-nearest neighbor coupling in addition to the Majorana
fermion analog of polyacetylene in the transverse ising model. The difference here is that onsite-potentials do not
occur with Majorana fermions: terms like $V a_{i}^{2}$ or $V b_{i}^{2}$ only add  overall constants, since $a_{i}^{2}=b_{i}^{2}=1$
We
write down the Hamiltonian in the basis $\psi^T=(a_1, b_1, a_2,
b_2, a_3, b_3...)$ and see that

\begin{equation}
H = -\frac{i}{2}\left( \begin{array}{ccccccccc}
0& 1 & 0& \ldots &\\
-1 & 0& \lambda_1& 0&\lambda_2 & \ldots &\\
0&
-\lambda_1&0&1&0&\ldots &\\
0&0&-1&0& \lambda_1&0&\lambda_2&\ldots &\\
\vdots & \vdots & & &\ddots
\end{array} \right).\label{HMajorana}
\end{equation}

The zero modes are given by the recursion relation of the amplitudes,
\begin{equation}
-A_i +\lambda_1 A_{i+1}+\lambda_2A_{i+2}=0,\label{eq:recurrence}
\end{equation}
where the eigenvector is chosen to be of the form $(A_{1}, 0, A_{2}, 0, \cdots)^{T}$. 
We note that when $\lambda_{2}< 0$, the above recursion relation looks like the equation of motion of a damped harmonic
oscillator with the time variable discretized.
The two linearly independent solutions can be expressed as:
\begin{equation}
A_i=C_1 q_{+}^{i}+C_2 q_{-}^{i},
\end{equation}
where $q_{\pm}$ satisfy
\begin{eqnarray}
1=\lambda_1 q+\lambda_2 q^2,\\
q_{\pm}=\frac{-\lambda_1\pm \sqrt{\lambda_1^2+4\lambda_2}}{2\lambda_2},
\label{recurrenceq}
\end{eqnarray}
and $C_{1}$, $C_{2}$ are  constants.
However, when we restrict ourselves to real and normalizable solutions, we may have only one, two, or  zero solutions.
Note that $\lambda_{1}^{2}+4\lambda_{2}>0$
is sufficient for giving us real solution regardless of $C_{1}$ to $C_{2}$ ratio.
Therefore, we can obtain the phase boundary for $\lambda_{1}^{2}+4\lambda_{2}>0$ just by examining whether $|q_{\pm}|$  is larger than 1 or
not. For the case $\lambda_{1}, \lambda_{2}>0$, we can recover the results found by Kopp and Chakravarty. For $1-\lambda_{1} < \lambda_{2}< 1+\lambda_{1}$,
we have $0< q_{+} <1<-q_{-}$, and therefore  a single Majorana zero mode at each end of the chain. For
$\lambda_{2}>1+\lambda_{1}$, we have $0< - q_{-} <1$ and $0<q_{+}<1$ and thus there are two Majorana zero modes at each end of the
chain. Lastly, we have no Majorana zero mode for $\lambda_{2}<1-\lambda_{1}$, as $q_{+}$ and $-q_{-}$ are both greater than 1.
We can extend this analysis to $\lambda_{1}^{2}+4\lambda_{2}>0$  and $\lambda_{2}<0$, $\lambda_{1}>0$. For $\lambda_{2} < 1-\lambda_{1}$ and $\lambda_{2}>-1$, we have $|q_{\pm}|>1$ and thus no
Majorana zero mode, whereas $\lambda_{2} > 1-\lambda_{1}$ gives us $0<q_{+}<1<-q_{-}$ and thus there is one Majorana zero mode at
each end of the chain.

By contrast, when we break the time reversal symmetry by having phase difference between the two pairing terms as in Eq.~(\ref{eq:complexBdG}), we find that we can have only zero or one normalizable Majorana zero mode at the end of the chain, as shown above from the explicit solution of the BdG equation. 
Interestingly, we find that we obtain the same result if we break time reversal symmetry by adding an impurity term of the form (see Appendix~\ref{imp-recur})
\begin{equation}
\begin{split}
H_{\text{imp}} &= -i\tilde{\lambda} a_j a_{j+m} \\ 
&= -i\tilde{\lambda}(c^\dagger_j c_{j+m} - c^\dagger_{j+m} c_j + c^\dagger_j c^\dagger_{j+m} - c_{j+m} c_j),
\end{split}
\label{eq:Himp}
\end{equation}
to the original Hamiltonian Eq.~\eqref{eq:Ham1}. Since the translational invariance is broken in this case, the number of zero Majorana modes provide a convenient way to distinguish different phases.

\subsection{Oscillatory Majorana zero modes with varying wavelength}
 We find from
the recursion relation, Eq.~(\ref{eq:recurrence}),  that there are no Majorana zero modes for $-1 < \lambda_{2} < 0$ and $\lambda_{2}<1-\lambda_{1}$. However, there are two oscillatory zero modes
for $\lambda_{2}< -1$ and $\lambda_{1}^{2}+4\lambda_{2}<0$, with amplitudes at a lattice site $j$  given by the two solutions of the recursion relation:
\begin{equation}
A_{j}= (-\lambda_{2})^{-j/2} \cos j\theta,
\end{equation}
and
\begin{equation}
A_{j}= (-\lambda_{2})^{-j/2} \sin j\theta,
\end{equation}
where $\theta = \arcsin(\lambda_1/\sqrt{-4\lambda_{2}})$. The amplitude could be rewritten as 
\begin{equation}
(-\lambda_{2})^{-j/2} = e^{-x/\xi} ,
\end{equation}
where $\xi= 2 a/\ln(-\lambda_{2})$. Note that close to the quantum critical line $|\lambda_{2}|\sim |\lambda_{1} -1|$. We have reintroduced the lattice spacing $a$ here. An example of oscillatory Majorana modes are shown in Fig.~(\ref{fig:fig6}).

When $\lambda_2$ becomes negative, an  oscillatory phase, as determined from the spin-spin correlation function,   was obtained from
a dual transformation that exchanges sites and bonds of the lattice. Then the  Hamiltonian in Eq.~(\ref{eq:Ham1}) can be  cast in the standard notation of the quantum $XY$-model  by factoring out an overall scale. Thus, with $\mu$'s as bond-centered Pauli matrices,
\begin{equation}
\begin{split}
H&=-\frac{2}{1+r}\sum_n\bigg[\frac{1+r}{2}\mu_1(n)\mu_1(n+1)\\&+\frac{1-r}{2}
\mu_2(n)\mu_2(n+1)+h\mu_3(n))\bigg],\label{Hxy}
\end{split}
\end{equation}
the two  parametrizations are related to each other by
\begin{equation}
\lambda_1 = \frac{2h}{1+r}, \;
\lambda_2 = \frac{r-1}{1+r}.
\end{equation}
The critical line in $XY$-model, separating the quantum disordered phase from the ferromagnetic phase, is $h=1$, which corresponds to
$\lambda_1+\lambda_2=1$,
separating the  ordered phase  from the disordered phase. The model was previously studied only in the range $0<r <1$ and $h>0$.
Since the ordered and the disordered phases are exchanged under duality,    the disordered phase of
the three-spin model is $\lambda_1+\lambda_2<1$.

A complex calculation~\cite{Barouch:1971} of the instantaneous spin-spin correlation function
showed  that within the ferromagnetic phase, there is an oscillatory   phase in the which the 
connected correlation function $G(x)= \langle \mu_{3}(x)\mu(0) \rangle -\langle \mu_{3}(x) \rangle \langle \mu_{3}(0) \rangle$
in the limit $x\to \infty$, is 
\begin{equation}
G(r) = \begin{cases} \frac{1}{\sqrt{x}}e^{-x/\xi}, &\text{disordered,}\\
                                     \frac{1}{x^{2}} e^{-x/\xi}, &\text{ordered,}\\
                                     \frac{1}{x^{2}}  e^{-2x/\xi} \Re(B e^{i K x}), &\text{oscillatory ordered.}
            \end{cases}
\end{equation}
Here $\cos K= \lambda_{1}/\sqrt{-4\lambda_{2}}$ and $\xi$ is the spin-spin correlation length. The oscillatory phase in the $XY$-model
is bounded by $r^2+h^2\leq 1$, which corresponds to 
$\lambda_2\leq-\lambda_1^2/4$ in the three-spin model. Note that the oscillation wavelength is identical to the wavelength of the Majorana fermions.
Even the correlation length close to criticality is the scale of the  exponential decay of the Majorana fermions. This must imply that in the spectral
decomposition, the Majorana zero modes asymptotically dominate, although we have not yet found a rigorous proof of it.

Since Majorana  modes of zero energy are degenerate eigenstates, a change of the number of Majorana zero modes can only occur when the energy gap collapses, i.e. at a quantum phase transition. 
Reexamining the behavior of Majorana zero mode in the full parameter space,  there are three  dividing lines based on their number: $\lambda_2=1+\lambda_1$, $\lambda_2=1-\lambda_1$ and $\lambda_2=-1$; see Fig~\ref{fig:fig1}.
These lines are identified with the critical lines signifying  phase transition, as can be seen in the energy spectrum in the previous result. So in this case, the number of Majorana zero modes serve as an ``order parameter'' for the quantum phase transition. 

We can then distinguish  phases and locate quantum phase transitions by simply transforming the
Hamiltonian in terms of  Majorana operators and finding
the number of allowed Majorana zero-modes at each end of the chain,
which is a topologically protected quantity. This provides us a profoundly simple way to study quantum phase transitions.

\subsection{Unbroken unitary time reversal symmetry}
\begin{figure}
\includegraphics[width=\linewidth]{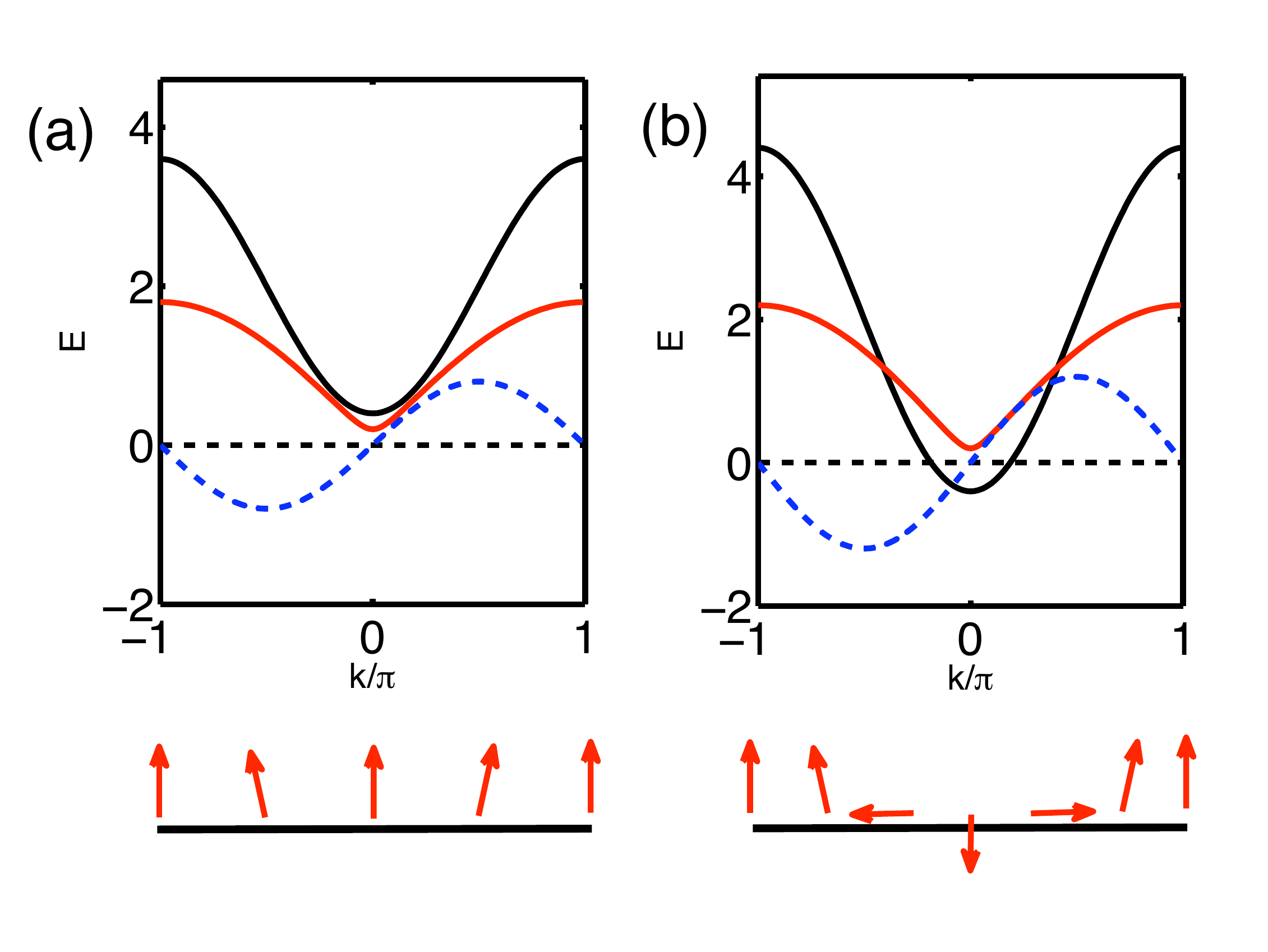}
\caption{(Color online)Topological phase transition across the point $\lambda_1 = 1, \lambda_2 = 0$.  (a) $\lambda_1 = 1^-, \lambda_2 = 0$.  The quantities plotted are $\epsilon_k$ (solid black line), $\mu$ (dashed black line), $\Delta(k)$ (dashed blue line), and quasiparticle energy (solid red line) as a function of momentum in the one dimensional Brillouin zone.   (b) The same quantities are plotted for $\lambda_1 = 1^+, \lambda_2 = 0$.  The associated Anderson pseudospin vector $\hat d(k) $ is drawn schematically below each plot.  It is clear that in (a) the pseudospin does not wind along the 1d Brillouin zone, i.e. $W = 0$,  whereas in (b) it winds once, i.e. $W = 1$.      }
\label{fig1}
\end{figure}

Some insight into the phase diagram can be obtained from the perspective of the weak to strong pairing topological phase transitions in this model.  Starting from the spinless Fermion Hamiltonian in Eq.~(\ref{eq:Ham1}), we get upon Fourier transformation
\begin{equation}
\begin{split}
H &= \sum_k \left(2 - 2 \lambda_1 \cos{k} - 2 \lambda_2 \cos{2 k} \right) c^{\dagger}_k c_k \\
&+ \sum_k \left( i \lambda_1 \sin{k} c^{\dagger}_{k} c^{\dagger}_{-k} + i \lambda_2 \sin{2 k} c^{\dagger}_{k} c^{\dagger}_{-k} + h.c. \right),
\end{split}
\end{equation}
which describes a superconductor  with a pairing potential that consists of a nearest neighbor and a second nearest neighbor  $p$-wave pairing.  The BdG Hamiltonian which governs the dynamics of the BCS quasiparticles at each momentum $k$ has the form 
\begin{eqnarray}
H_{BdG}(k) = \left( \begin{array}{cc} \epsilon_k - \mu & i \Delta(k) \\ -i \Delta(k) & \mu - \epsilon_k \end{array} \right)
\end{eqnarray}
where $\epsilon_k = -2 \lambda_1 \cos{k} -2 \lambda_2 \cos{2 k}$, $\Delta(k) = \lambda_1 \sin{k} + \lambda_2 \sin{2 k}$, and $\mu = -2$.  
In this representation of the model, the various phase boundaries described in previous sections correspond to Lifshitz transitions, across which the number of Fermi points  change.  However, as we shall see, {\it not all Lifshitz transitions are topological phase transitions}.  To determine precisely whether a Lifshitz transition is a topological phase transition, we must define an integer-valued topological invariant that changes only across a topological phase transition.  It is convenient to define the invariant using the  Anderson pseudospin vector~\cite{Anderson:1958}
\begin{equation}
\vec d(k)  = \Delta(k) \hat y + \left( \epsilon_k - \mu \right) \hat z.
\end{equation}
In terms of this vector $H_{BdG}(k) = \vec d(k) \cdot \vec \tau$, where $\vec \tau$ are Pauli matrices which act in the Nambu (i.e. particle-hole) basis of $H_{BdG}$.  It is important to highlight that the pseudospin is defined only in the $yz$ plane in this problem.  This is a consequence of time-reversal symmetry (applied to spinless Fermions, time-reversal is simply the operation of complex-conjugation).  Time-reversal symmetry ensures that the {\it relative} phase between the nearest neighbor and second neighbor pairing amplitudes must be real (in the context of the spin model, the relative phase is identically zero).  However, when time-reversal symmetry is broken, the relative phase between these can be an arbitrary complex number.  If this happens, the Anderson pseudospin vector will have {\it three} components, and the analysis below is invalidated.  In this section, we shall restrict our attention to  the case where the relative phase is zero. The topological invariant that characterizes the phase transitions  will be defined in terms of the unit vector 
\begin{equation}
\hat d(k) = \frac{\vec d(k)}{\vert \vec d(k) \vert}   \equiv \cos{\theta_k} \hat y + \sin{\theta_k} \hat z .
\end{equation}  
Here, the momentum states with periodic boundary conditions form a ring $T^1$, and the unit vector $\hat d(k)$, lives on a unit circle $S^1$ in the $yz-$plane.  Therefore, the angle $\theta(k)$ is a mapping $\theta(k): S^1 \rightarrow T^1$ and the topological invariant we seek is simply the fundamental group of this mapping, which is just the integer {\it winding number}
\begin{equation}
W =  \oint \frac{d \theta_k}{2 \pi}
\end{equation}
where the integral is done around the one dimensional Brillouin zone.  This quantity characterizes the number of times the vector $\hat d(k)$ rotates in the $yz$-plane around the one-dimensional Brillouin zone.  It can only be an integer and therefore cannot vary with smooth deformations of the Hamiltonian, so long as the quasiparticle gap remains finite.  The winding number changes discontinuously only when the energy gap vanishes, i.e. at a topological phase transition.  
Moreover, {\it the change in the number of normalizable Majorana modes at each end of the chain across a transition is given by the change in the winding number} $W$\cite{Read:2000}.  
We now apply this framework to characterize several critical points in the $\lambda_1 \lambda_2$ plane.  In Fig.~\ref{fig1} we show the results  in the vicinity of  the critical point at $\lambda_1 = 1$ and  $\lambda_2 = 0$.  At this critical point, the chain consists only of nearest neighbor hopping and pairing.  Therefore,  the kinetic energy has its minimum at $k=0$, and the gap function also vanishes at this point.  For $\lambda_1 = 1^{-}$ and $\lambda_2 = 0$, the chemical potential occurs below the band bottom.  In this limit, the winding number of $\hat d(k)$ is zero, since the configuration is topologically equivalent to one where $\theta(k) = \pi/2$ for all $k$, and this state clearly has $W=0$.  On the other side of the transition, $\lambda_1 = 1^+$ and  $\lambda_2 = 0$,  the chemical potential crosses the band bottom, and the winding number changes to $W = 1$.  At the critical point itself, the Anderson pseudospin unit-vector is not defined at the point $k=0$, where the gap closure occurs.  The change in $W$ is identical to the change in the number of normalizable Majorana fermion zero modes  across this transition.  

Next, we use similar reasoning to study the transition at the critical point $\lambda_2 = 1$ and  $\lambda_1 = 0$ across which the {\it change} in the number of normalizable Majorana modes at each end is 2.  Here, the chain consists only of second-neighbor hopping and pairing.  Note that there are now two extremal points of the bandstructure: one at $k=0$, the other at $k=\pi$.  The winding number jumps from 0 to 2 across this transition, and  for $\lambda_2 > 1$ and  $\lambda_1 = 0$, we see 2 normalizable Majorana zero energy modes at each end of the chain.  
\begin{figure}
\includegraphics[width=\linewidth]{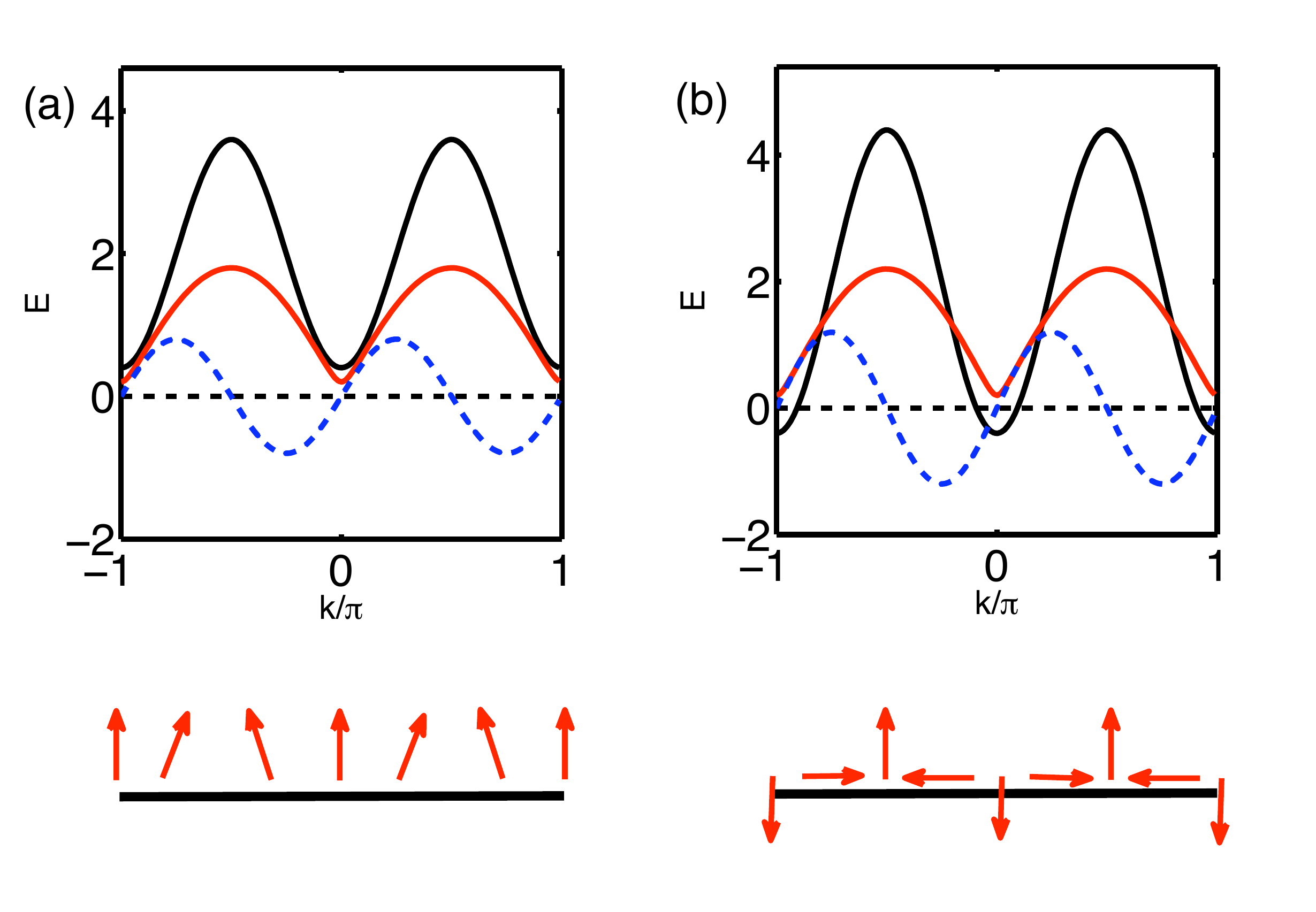}
\caption{(Color online)Topological phase transition in the vicinity of the point $\lambda_1 = 0, \lambda_2 = 1$.  (a) $\lambda_2 = 1^-, \lambda_1 = 0$; (b)  $\lambda_2 = 1^+, \lambda_1 = 0$.    It is clear that in (a), the pseudospin does not wind along the 1d Brillouin zone, i.e. $W = 0$,  whereas in (b) it winds twice, i.e. $W = 2$.      }
\label{fig2}
\end{figure}

Interestingly, our model hosts both BEC-BCS {\it transitions} and BEC-BCS {\it crossovers}.  Only the former are topological transitions:  these require that  (1) an extremum of the band crosses the chemical potential, and (2) the pairing potential vanishes at the same momentum.  If an extremum of the band crosses the chemical potential at a point where the gap does {\it not} vanish, the winding number will {\it not} change, since the total energy gap does not vanish.  This is an example of a BCS-BEC {\it crossover}, and not a {\it transition}.  This type of crossover is seen near the line $\lambda_2  = -1$.   
\begin{figure}
\includegraphics[width=\linewidth]{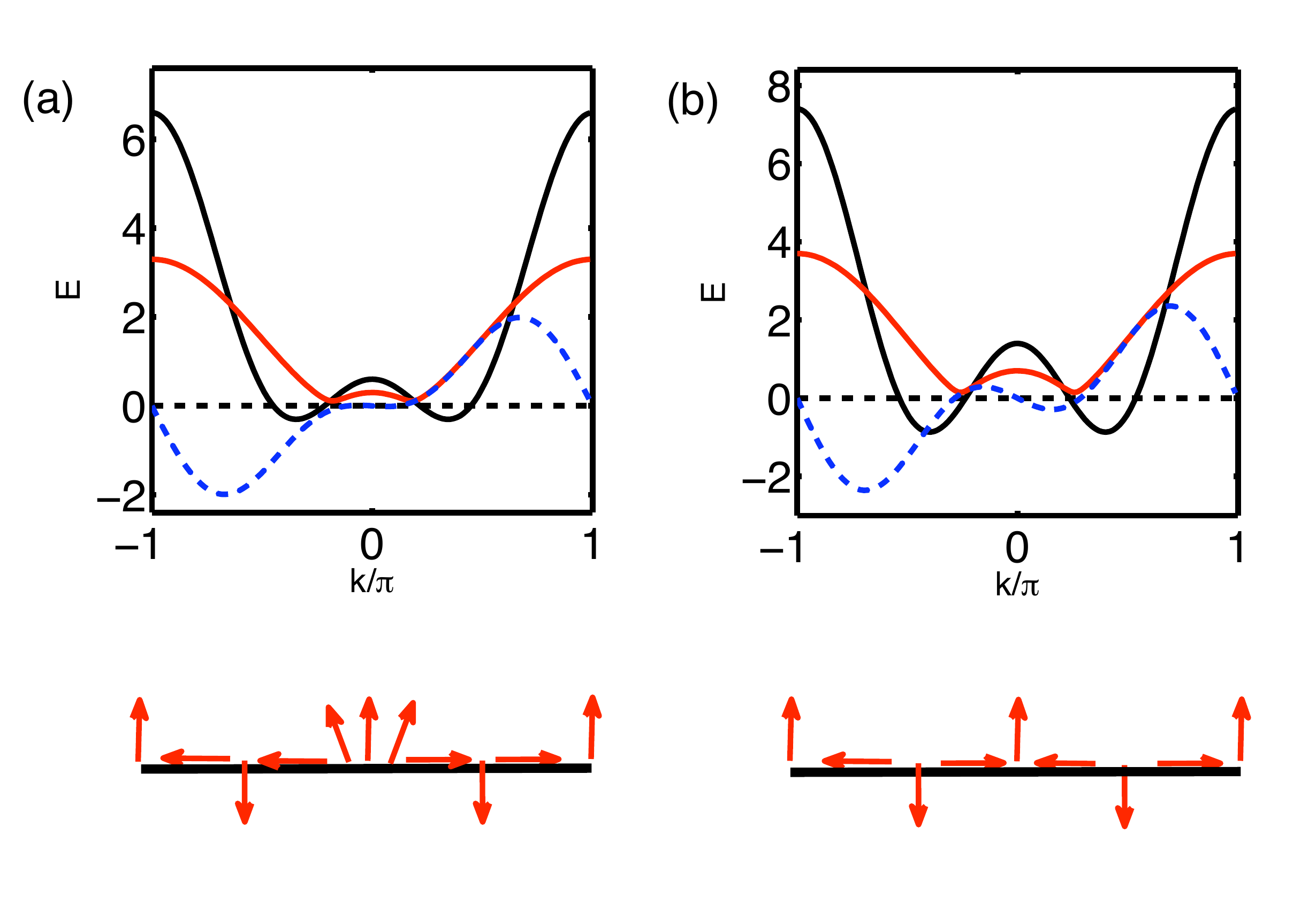}
\caption{(Color online)Topological phase transition in the vicinity of the point $\lambda_1 = 1.5, \lambda_2 = -1$.  (a) $\lambda_2 = -1^+, \lambda_1 = 1.5$;  (b)  $\lambda_2 = -1^-, \lambda_1 = 1.5$.  In (a), the pseudospin does not wind along the 1d Brillouin zone, i.e. $W = 0$,  whereas in (b) it winds twice, i.e. $W = 2$.      }
\label{fig3}
\end{figure}
An illustrative example is presented in Fig. \ref{fig3}.  Here, the critical point $\lambda_1 = 1.5$ and $\lambda_2 = -1$ is studied.  In Fig. \ref{fig3}a, the properties of the system are shown at $\lambda_1 = 1.5$ and  $\lambda_2 = -1^+$, where no normalizable Majorana  zero mode occurs at the boundary.  From the fact that Fermi points occur in this system, it is clear that the system is in the BCS regime.  However, it is apparent from the form of the Anderson pseudospin that the winding number is identically zero.  Thus, while this state is a BCS state, it is topologically equivalent to a BEC state which also has zero winding number.  Thus, a {\it crossover} can connect this state to a BEC state.  However, when $\lambda_1 = 1.5$ and $\lambda_2 = -1^-$, i.e. just below the critical point, we know from the analysis of previous sections that there are 2 normalizable Majorana fermion zero modes at each edge of the chain.  This  is also consistent with the winding number of the Anderson pseudospin, which is $W = 2$ in this regime.  
We stress therefore, that a topological phase transition between 2 BCS states can occur.  However so long as the bandstructure possesses inversion symmetry, it follows that such topological BCS-BCS transitions can only change the topological invariant by $\pm 2$.    In a similar way, the critical line in Fig.~(\ref{fig:fig1}) from b to d represents a topological transition across which the winding number changes by 2.  Along this line, the gap vanishes at an incommensurate set of points in momentum space.   
\begin{figure}
\includegraphics[width=\linewidth]{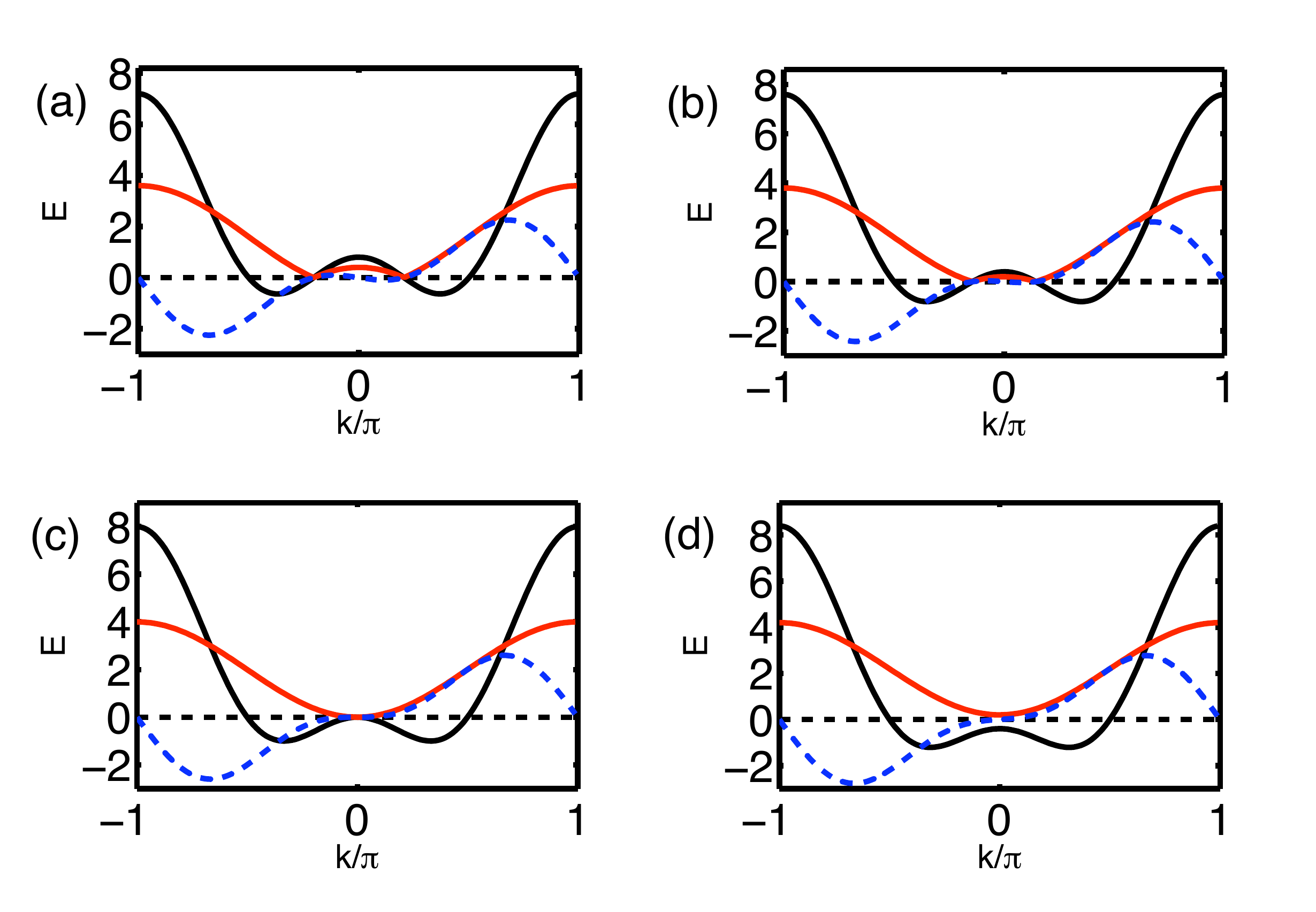}
\caption{(Color online)Topological phase transition along the line $\lambda_2 = -1$.  (a) $\lambda_1 = 1.6, \lambda_2 = -1$.  (b) $\lambda_1 = 1.8, \lambda_2 = -1$. (c) $\lambda_1 = 2.0, \lambda_2 = -1$.  (d) $\lambda_1 = 2.2, \lambda_2 = -1$.   The associated Anderson pseudospin vector is not shown since it is not defined at points where the quasiparticle energy gap vanishes.  However, the system in (d) is fully gapped and has $W=1$, which is consistent with the analysis of previous sections.      }
\label{fig4}
\end{figure}
In (c), the point $\lambda_1 = 1$ and $\lambda_2 = -1$ is considered.   Here again, there are two band minima.  However, they occur at incommensurate momenta, $\pm k_0$, symmetric about the origin.  Therefore this critical point marks a transition from 0 to 2 Majorana zero modes at each end of the chain.  This is the point (c)  in Fig.~(\ref{fig:fig1}).  As we approach the special multicritical point d in Fig.~(\ref{fig:fig1}) ($\lambda_1 = 2, \lambda_2 = -1)$, these two incommensurate points move towards the origin.  They meet at $k=0$, which is now a local {\it maximum} of the bandstructure.  The 2 momenta $\pm k_0$ meet at $k=0$ at the multicritical point ($\lambda_1 = 2, \lambda_2 = -1$).  In this way, all the topological phase transitions that occur in the model can be understood.

Lastly, we study the nature of the special multi-critical point which occurs at ($\lambda_1 =2, \lambda_2 = -1$).  As this multicritical point is approached, the two incommensurate momenta which occur around on either sides of $k=0$, where the gap vanishes, approach each other at $k=0$.  At the multicritical point, the momenta meet at $k=0$ and annihilate each other, as shown in Fig. \ref{fig4}.   Note that in Fig. \ref{fig4} (a-c), the Anderson pseudospin is not defined at points where the gap vanishes.  Therefore, the winding number itself is not defined (this is consistent with the fact that this line represents a topological phase transition).  However, for $\lambda_1 > 2, \lambda_2 = -1$, the system is gapped everywhere, has a winding number $W=1$, and possess one normalizable Majorana fermion zero mode at each end of the chain. 

To conclude, in this section, we have described a complementary way in which the phase transitions in this model can be understood.  Specifically, we have introduced the topological invariant corresponding to the winding number of the Anderson pseudospin vector around the Brillouin zone.  
Each of the phase boundaries of the spin model studied in this paper corresponds to regions where the winding  number is ill defined, and across each critical point, the winding number changes by an integer.  The number of normalizable Majorana fermion zero energy modes localized to each end of the chain at a point in the $\lambda_1 \lambda_2$-plane is exactly equal to the winding number at that point.  We have also emphasized the point that a crossover can occur between a BEC and BCS system provided that both have the same winding number ($ W = 0$), and we have also demonstrated that there can be topological phase transitions from one type of BCS state to another.

\subsection{Broken time reversal symmetry}
 
When the phase difference between the nearest neighbor and the second neighbor pairing amplitudes is non-zero, the BdG Hamiltonian takes  the form 
\begin{equation}
H(k) 
 = \left( 
\begin{array}{cc}
\xi_k & -\alpha_k +i \beta_k \\
-\alpha_k -i \beta_k & -\xi_k
\end{array}
\right),
\label{H:BdG}
\end{equation}
where $\alpha_k = \lambda_2 \sin \theta \sin 2k$ and $\beta_k = \lambda_1 \sin k + \lambda_2 \cos \theta \sin 2k$. The particle-hole symmetry of the BdG Hamiltonian is
\begin{equation}
\sigma_{1} H(k) \sigma_{1}= - H(-k)^{*}.
\end{equation}
The $k=0,\pm \pi$ are special because they map onto themselves. Then  from the equations
\begin{eqnarray}
H(k=0) &=& (1 - \lambda_1 - \lambda_2)  \sigma_3, \\
H(k=\pi) &=& (1 +\lambda_1 - \lambda_2)  \sigma_3,
\end{eqnarray}
 it follows that the topological invariant is $\prod_{k=0,\pi} {\rm sgn}(\xi_k)$.  When  $\xi_{0}=(1 - \lambda_1 - \lambda_2) $ and $\xi_{\pi}=(1 + \lambda_1 - \lambda_2)$ have opposite signs, we get $n=1$, otherwise $n=0$. As to the physical significance of $k=0,\pm \pi$, it is  similar to the case of topological insulators~\cite{Hasan:2011,*Qi:2010} where these points are  termed ``time reversal invariant" points. In a superconductor, $k$ and -$k$ states are paired, so the $k=0, \pm \pi$ points are again special because they map onto themselves. This elegant topological  argument  due to R. Roy~\cite{[{Personal communication.\;}]Roy:2011} confirms the results in Sec.~\ref{sec:secIIIB}; see also Refs. ~\onlinecite{Ghosh:2010,Kitaev:2001}.

\section{Conclusions}
In this paper we have studied an exactly solvable spin Hamiltonian that is TFIM with an added  three spin interaction.  While such a spin interaction may appear to be artificial to the reader, such a term is generated in real-space renormalization group treatments of TFIM.~\cite{Hirsch:1979}  Therefore, it is a physically plausible coupling in a more realistic Hamiltonian and corresponds to longer ranged pairing and hybridization interactions among the fermions related to the spins via a Jordan-Wigner transformation.  By analyzing the fermionized version of this spin Hamiltonian we have identified the quantum phase transitions in this system and to classify them according to the number of  Majorana zero modes localized at each end.  This number in turn is related to the winding number of the Anderson Pseudospin unit vector along the one dimensional Brillouin zone, so long as time-reversal symmetry for the spinless fermions is preserved.  We have noted that when there are an even number of Majorana fermion modes at each end, there can be a crossover from a regime where the Majorana fermion wave-function decays in an oscillatory fashion (with an exponential envelope) to a regime where these modes decay exponentially without oscillation.  Interestingly, at the crossover, the entanglement entropy vanishes identically.  We stress that this crossover does not occur when there are an odd number of Majorana fermions at each end.  Whether the vanishing of the entanglement entropy is a necessary condition for this crossover remains to be understood.  The degree to which such a crossover remains generic, or is ascribed to the integrability of the spin chain is unclear.  Finally, an interesting possibility is that such crossovers may occur in higher dimensions in spin triplet superconductors in the presence of vortices, and other topological superconductors involving non-centrosymmetric systems.  We shall relegate these studies to future work..

For unitary time reversal invariance, the topological argument involving Anderson's pseudospin vector leads to  winding number $\mathbb{Z}$. One might wonder if higher windings beyond $n=0, 1, 2$ are possible as well. In principle, it is. To check, we added an even longer ranged term $H_{3}=\lambda_{3}c_{i}^{\dagger}c_{i+3}+ \lambda_{3} c_{i}^{\dagger}c_{i+3}^{\dagger}+ h.c.$ Now, in addition to $n=0,1,2$, we also get winding number  $n=3$ in  appropriate regimes of the parameter space from explicit calculations of the BdG equation. It is quite likely that these higher order windings are energetically punished. The situation is very similar to the  $XY$-model in two dimensions  for which higher order vorticity is suppressed by the chemical potential. Clearly phases with $n=\mathbb{Z}$ Majorana zero modes are  allowed by  longer ranged Hamiltonians. We find this phenomena intriguing, which deserves further attention. However, once the protection due to unitary time reversal invariance is removed the topological invariant collapses to $\mathbb{Z}_{2}$, with at most one Majorana zero mode at each end of an open chain.

We have previously emphasized that while the solution of the spin model is exact, the fermonized version is  a mean-field description of a $p$-wave superconductor whose exact solution requires treatment of fluctuation effects. In a recent paper it has been shown, however, that  including fluctuation effects do not change the basic picture in a one-dimensional model.~\cite{Fidkowski:2011,*Sau:2011} Whether such a conclusion holds in higher dimensions, where Majorana zero modes are nucleated in the vortex cores of a $p_{x}+ip_{y}$ superconductors, remains to be seen. We leave this problem for future  research.

 An interesting question is  whether or not the topological  phases described here are perturbatively stable against weak interactions.  We believe that they  are,  because they are gapped.  In principle, for stronger interactions, the unitary time-reversal symmetry that protects the $n=2$  phase can break spontaneously and destabilize it. The effect of stronger interactions in  a specific model  has been considered in Ref.~\onlinecite{Fidkowski:2010}.

\begin{acknowledgments}
 Yuezhen Niu thanks UCLA physics department for its
hospitality. S. C. and Y. N. were supported by US NSF under the Grant DMR-1004520. S.B.C was supported by the DOE under contract DE-AC02-76SF00515. I. M. was funded by funds from the David S. Saxon Presidential Chair at UCLA. S.R. is supported by startup funds at Stanford University. We thank Parsa Bonderson, Cristina Bena, Pallab Goswami, Alexei Kitaev, Roman Lutchyn, Chetan Nayak, Rahul Roy,   Kirill Shtengel and  Matthias Troyer for comments.
 This work was partly carried out at the Aspen Center for Physics.
\end{acknowledgments}

\appendix
\section{Broken time reversal invariance}
\begin{widetext}
When there is a relative phase $e^{i\theta}$ between the nearest neighbor and the next-nearest neighbor pairing amplitudes the Majorana Hamiltonian is
\begin{eqnarray}
\mathcal{H} &=&
-i \{ -\sum^{N}_{i=1} b_{i} a_{i} + \lambda_1 \sum^{N-1}_{i=1} b_{i} a_{i+1} 
+ \frac{\lambda_2}{2} \sum^{N-1}_{i=2} [ (1+\cos \theta ) b_{i-1} a_{i+1} \nonumber \\
&& \hspace{0.5in} -(1-\cos \theta ) a_{i-1} b_{i+1} + \sin \theta (a_{i-1} a_{i+1} -  b_{i-1} b_{i+1}) ] \}. 
\label{H2}
\end{eqnarray}
Thus, we cannot simply set $\lambda_{2}$ to be complex in Eq.~(\ref{eq:recurrence}).
The  Hamiltonian in the Majorana basis $\psi^{T}=(a_1,b_1,a_2,b_2,a_3,b_3, \cdots )$ is
\begin{equation} \label{H3}
H= -\frac{i}{2} \left( 
\begin{array}{cccccccc}
0 & 1 & 0 & 0 & \frac{\lambda_2}{2}\sin \theta & -\frac{\lambda_2}{2}(1-\cos \theta)  & \cdots & \cdots \\
-1 & 0 & \lambda_1 & 0 & \frac{\lambda_2}{2}(1+\cos \theta) & -\frac{\lambda_2}{2}\sin \theta & \cdots & \cdots \\
0 & -\lambda_1 & 0 & 1 & 0 & 0  & \cdots & \cdots \\
0 & 0 & -1 & 0 & \lambda_1 & 0  & \cdots & \cdots \\
\vdots & \vdots & 0 & -\lambda_1 & 0 & 1 & \cdots & \cdots \\
\vdots & \vdots & 0 & 0 & -1 & 0 & \cdots & \cdots \\
\vdots & \vdots & \vdots & \vdots & \vdots & \ddots & \ddots & \cdots \\
\end{array}
\right).
\end{equation}
To find the zero mode eigenvectors, we can try a solution of the form: $|\Psi \rangle =(A_1,B_1,A_2,B_2,A_3,B_3, \cdots )^{T}$. However, the recursion relations turn out to be 
too complex to solve  analytically. Thus, we resorted to  numerical 
diagonalization of the BdG Hamiltonian in the main text.
\end{widetext}

\section{Majorana zero modes in the presence of impurity}
\label{imp-recur}
In general, we find that when $H$ of Eq.~\eqref{eq:Ham1} results  in two Majorana zero mode, $H_{imp}$ in Eq.~(\ref{eq:Himp}) destroys them (except for some special cases), while the  regime with one Majorana zero mode remains intact. 

Consider the general definition of the Majorana zero mode $\Gamma = \sum(A_i a_i + B_i b_i)$, which is determined by the commutator
\begin{equation}
\begin{split}
0 = & [H_0 + H_{\text{imp}}, \Gamma] \\ =&  2i \sum (A_i - \lambda_1 A_{i+1} - \lambda_2 A_{i+2}) b_i\\
-& 2i B_1 a_1 -2i(B_2 - \lambda_1 B_1)a_2\\ -&2i\sum (B_{i+2} -\lambda_1 B_{i+1} - \lambda_2 B_i)a_{i+2} \\ + &2i\tilde{\lambda}(A_{j+m} a_j - A_j a_{j+m}),
\end{split}
\end{equation}
which requires, in addition to the original recursion formula
\begin{eqnarray}
&A_i - \lambda_1 A_{i+1} - \lambda_2 A_{i+2} = 0,\\
&B_{i} -\lambda_1 B_{i-1} - \lambda_2 B_{i-2} 
=0, (i>j, \,i\neq j+m),
\end{eqnarray}
new boundary conditions for $B_j$'s:
\begin{eqnarray}
&B_i = 0\,\,\,{\rm for}\,\,\,i < j,\\
&B_j = \tilde{\lambda} A_{j+m},\\
&B_{j+m} - \lambda_1 B_{j+m-1} - \lambda_2 B_{j+m-2} 
= -\tilde{\lambda} A_j.\\
\end{eqnarray}
(note that the $A_i$ recursion relation is not affected by $B_i$'s). 
Because of this change in the boundary conditions, we can no longer set $B_i = 0$ for all $i$. 
Rather, for $i > j+m$, 
the general solution for$A_{i}$ and $B_{i}$ are of the form
\begin{align}
A_i =& C_+ q_+^i  + C_- q_-^i,\nonumber\\
B_{i} =& C'_+ (1/q_+)^i + C'_- (1/q_-)^i,
\label{genForm}
\end{align}
where $1 - \lambda_1 q_\pm - \lambda_2 q_\pm^2 = 0$. 

We can now see how the impurity term Eq.~(\ref{eq:Himp}) may destroy the Majorana zero modes. Eq.~\eqref{genForm} implies that if we had two Majorana zero modes without the impurity, which only requires $|q_\pm|<1$, we will not have any 
normalizable Majorana zero mode due to the divergence of $B_i$ unless we have $B_{j+m-1} = B_{j+m} =0$, which  can occur only under special situations. On the other hand, if we had a single Majorana zero mode without the impurity, which means $|q_+|<1<|q_-|$, the Majorana zero mode survives if  $B_{i+m}/B_{i+m-1}=1/q_{-}$. We have checked this explicitly  for the special cases of $m=1,2,3$.

\section{Impurity induced tunneling between two Majorana zero modes}
\label{impTunnel}

To consider the condition for the stability of two Majorana zero mode, we first not that, in the limit where the bulk gap is large, a semi-infinite chain can be regarded as a two-state system. This is because the two Majorana zero mode would form a single zero energy state, giving us energy degeneracy between the case where this zero energy state is occupied and the case where this zero energy state is vacant. 
Due to the fermion number parity conservation, perturbation cannot give rise to any off-diagonal term 
between the two states; all we can obtain is the energy difference between the occupied and vacant zero energy state.

Therefore, an impurity term can annihilate the two Majorana zero modes if the mode expansion of this impurity term gives rise to dependence on the occupancy of the zero energy state. We know that, in absence of any impurity, the Hamiltonian in Eq.~(\ref{eq:Ham1})
gives us two Majorana zero modes 
near $i=1$ can be written down as the linear combination of only $a_i$'s:
\begin{equation}
\Gamma_n = \sum_i c_{ni} a_i,
\end{equation}
where $n=1,2$ and $c_i \in \mathcal{R}$. 
Then, the annihilation operator of the zero energy state can be written as
\begin{equation}
f_0 = (\Gamma_1 + i\Gamma_2)/2 = \sum_i (c_{1i}+ic_{2i})a_i/2.
\label{EQ:zero}
\end{equation}
What follows from this is that when we do the mode expansion on Majorana fermions on each site, only $a_i$'s receive contribution from the zero energy state whereas all $b_i$'s do not:
\begin{align}
a_i =& (\tilde{c}_{i0} f_0 + \tilde{c}^*_{i0} f^\dagger_0) + \sum_{m=1}^\infty (\tilde{c}_{im} f_m+\tilde{c}^*_{im} f^\dagger_m),\nonumber\\
b_i =& \sum_{m=1}^\infty (\tilde{c}'_{im} f_m+\tilde{c}'^*_{im} f^\dagger_m).
\label{EQ:modeExpand}
\end{align}
Any additional fermionic bilinear terms to the Hamiltonian cannot affect the zero modes unless mode expansion of such terms  contain $f^\dagger_0 f_0$. 

We make a further restriction that we demand the fermionic bilinear to be local. The criterion for locality here is that, if our fermion operators are from sites $i, j$, they should satisfy $|i-j| \sim$ O(1).

This leads to the conclusion that 
only $i a_i a_j$ can gap out the zero modes, while $i a_i b_j$ and $i b_i b_j$ do not. (Conversely, 
if we had the right end of the semi-infinite chain, it is $i b_i b_j$ that gaps out the zero modes.) 
We see from Eq.\eqref{EQ:modeExpand} 
\begin{equation}
i a_i a_j = -i (\tilde{c}_{i0} \tilde{c}^*_{j0} - \tilde{c}^*_{i0} \tilde{c}_{j0}) f^\dagger_0 f_0 + ({\rm gapped})
\end{equation}
but the mode expansions of $i a_i b_j$ and $i b_i b_j$ do not have the $f^\dagger_0 f_0$ term.


\begin{thebibliography}{30}%
\makeatletter
\providecommand \@ifxundefined [1]{%
 \@ifx{#1\undefined}
}%
\providecommand \@ifnum [1]{%
 \ifnum #1\expandafter \@firstoftwo
 \else \expandafter \@secondoftwo
 \fi
}%
\providecommand \@ifx [1]{%
 \ifx #1\expandafter \@firstoftwo
 \else \expandafter \@secondoftwo
 \fi
}%
\providecommand \natexlab [1]{#1}%
\providecommand \enquote  [1]{``#1''}%
\providecommand \bibnamefont  [1]{#1}%
\providecommand \bibfnamefont [1]{#1}%
\providecommand \citenamefont [1]{#1}%
\providecommand \href@noop [0]{\@secondoftwo}%
\providecommand \href [0]{\begingroup \@sanitize@url \@href}%
\providecommand \@href[1]{\@@startlink{#1}\@@href}%
\providecommand \@@href[1]{\endgroup#1\@@endlink}%
\providecommand \@sanitize@url [0]{\catcode `\\12\catcode `\$12\catcode
  `\&12\catcode `\#12\catcode `\^12\catcode `\_12\catcode `\%12\relax}%
\providecommand \@@startlink[1]{}%
\providecommand \@@endlink[0]{}%
\providecommand \url  [0]{\begingroup\@sanitize@url \@url }%
\providecommand \@url [1]{\endgroup\@href {#1}{\urlprefix }}%
\providecommand \urlprefix  [0]{URL }%
\providecommand \Eprint [0]{\href }%
\providecommand \doibase [0]{http://dx.doi.org/}%
\providecommand \selectlanguage [0]{\@gobble}%
\providecommand \bibinfo  [0]{\@secondoftwo}%
\providecommand \bibfield  [0]{\@secondoftwo}%
\providecommand \translation [1]{[#1]}%
\providecommand \BibitemOpen [0]{}%
\providecommand \bibitemStop [0]{}%
\providecommand \bibitemNoStop [0]{.\EOS\space}%
\providecommand \EOS [0]{\spacefactor3000\relax}%
\providecommand \BibitemShut  [1]{\csname bibitem#1\endcsname}%
\let\auto@bib@innerbib\@empty
\bibitem [{\citenamefont {Wilczek}(2009)}]{wilczek:2009}%
  \BibitemOpen
  \bibfield  {author} {\bibinfo {author} {\bibfnamefont {F.}~\bibnamefont
  {Wilczek}},\ }\href@noop {} {\bibfield  {journal} {\bibinfo  {journal}
  {Nature Phys.}\ }\textbf {\bibinfo {volume} {5}},\ \bibinfo {pages} {614}
  (\bibinfo {year} {2009})}\BibitemShut {NoStop}%
\bibitem [{\citenamefont {Moore}\ and\ \citenamefont
  {Read}(1991)}]{Moore:1991}%
  \BibitemOpen
  \bibfield  {author} {\bibinfo {author} {\bibfnamefont {G.}~\bibnamefont
  {Moore}}\ and\ \bibinfo {author} {\bibfnamefont {N.}~\bibnamefont {Read}},\
  }\href@noop {} {\bibfield  {journal} {\bibinfo  {journal} {Nucl. Phys. B}\
  }\textbf {\bibinfo {volume} {360}},\ \bibinfo {pages} {362} (\bibinfo {year}
  {1991})}\BibitemShut {NoStop}%
\bibitem [{\citenamefont {Nayak}\ and\ \citenamefont
  {Wilczek}(1996)}]{Nayak:1996}%
  \BibitemOpen
  \bibfield  {author} {\bibinfo {author} {\bibfnamefont {C.}~\bibnamefont
  {Nayak}}\ and\ \bibinfo {author} {\bibfnamefont {F.}~\bibnamefont
  {Wilczek}},\ }\href@noop {} {\bibfield  {journal} {\bibinfo  {journal} {Nucl.
  Phys. B}\ }\textbf {\bibinfo {volume} {479}},\ \bibinfo {pages} {529}
  (\bibinfo {year} {1996})}\BibitemShut {NoStop}%
\bibitem [{\citenamefont {Ivanov}(2001)}]{Ivanov:2001}%
  \BibitemOpen
  \bibfield  {author} {\bibinfo {author} {\bibfnamefont {D.~A.}\ \bibnamefont
  {Ivanov}},\ }\href@noop {} {\bibfield  {journal} {\bibinfo  {journal} {Phys.
  Rev. Lett.}\ }\textbf {\bibinfo {volume} {86}},\ \bibinfo {pages} {268}
  (\bibinfo {year} {2001})}\BibitemShut {NoStop}%
\bibitem [{\citenamefont {Fu}\ and\ \citenamefont {Kane}(2008)}]{Fu:2008}%
  \BibitemOpen
  \bibfield  {author} {\bibinfo {author} {\bibfnamefont {L.}~\bibnamefont
  {Fu}}\ and\ \bibinfo {author} {\bibfnamefont {C.~L.}\ \bibnamefont {Kane}},\
  }\href@noop {} {\bibfield  {journal} {\bibinfo  {journal} {Phys. Rev. Lett.}\
  }\textbf {\bibinfo {volume} {100}},\ \bibinfo {pages} {096407} (\bibinfo
  {year} {2008})}\BibitemShut {NoStop}%
\bibitem [{\citenamefont {Sau}\ \emph {et~al.}(2010{\natexlab{a}})\citenamefont
  {Sau}, \citenamefont {Lutchyn}, \citenamefont {Tewari},\ and\ \citenamefont
  {Das~Sarma}}]{Sau:2010}%
  \BibitemOpen
  \bibfield  {author} {\bibinfo {author} {\bibfnamefont {J.~D.}\ \bibnamefont
  {Sau}}, \bibinfo {author} {\bibfnamefont {R.~M.}\ \bibnamefont {Lutchyn}},
  \bibinfo {author} {\bibfnamefont {S.}~\bibnamefont {Tewari}}, \ and\ \bibinfo
  {author} {\bibfnamefont {S.}~\bibnamefont {Das~Sarma}},\ }\href@noop {}
  {\bibfield  {journal} {\bibinfo  {journal} {Phys. Rev. Lett.}\ }\textbf
  {\bibinfo {volume} {104}},\ \bibinfo {pages} {040502} (\bibinfo {year}
  {2010}{\natexlab{a}})}\BibitemShut {NoStop}%
\bibitem [{\citenamefont {Sau}\ \emph {et~al.}(2010{\natexlab{b}})\citenamefont
  {Sau}, \citenamefont {Tewari}, \citenamefont {Lutchyn}, \citenamefont
  {Stanescu},\ and\ \citenamefont {Das~Sarma}}]{Sau:2010b}%
  \BibitemOpen
  \bibfield  {author} {\bibinfo {author} {\bibfnamefont {J.~D.}\ \bibnamefont
  {Sau}}, \bibinfo {author} {\bibfnamefont {S.}~\bibnamefont {Tewari}},
  \bibinfo {author} {\bibfnamefont {R.~M.}\ \bibnamefont {Lutchyn}}, \bibinfo
  {author} {\bibfnamefont {T.~D.}\ \bibnamefont {Stanescu}}, \ and\ \bibinfo
  {author} {\bibfnamefont {S.}~\bibnamefont {Das~Sarma}},\ }\href@noop {}
  {\bibfield  {journal} {\bibinfo  {journal} {Phys. Rev. B}\ }\textbf {\bibinfo
  {volume} {82}},\ \bibinfo {pages} {214509} (\bibinfo {year}
  {2010}{\natexlab{b}})}\BibitemShut {NoStop}%
\bibitem [{\citenamefont {Alicea}(2010)}]{Alicea:2010}%
  \BibitemOpen
  \bibfield  {author} {\bibinfo {author} {\bibfnamefont {J.}~\bibnamefont
  {Alicea}},\ }\href@noop {} {\bibfield  {journal} {\bibinfo  {journal} {Phys.
  Rev. B}\ }\textbf {\bibinfo {volume} {81}},\ \bibinfo {pages} {125318}
  (\bibinfo {year} {2010})}\BibitemShut {NoStop}%
\bibitem [{\citenamefont {Lutchyn}\ \emph {et~al.}(2010)\citenamefont
  {Lutchyn}, \citenamefont {Sau},\ and\ \citenamefont
  {Das~Sarma}}]{Lutchyn:2010}%
  \BibitemOpen
  \bibfield  {author} {\bibinfo {author} {\bibfnamefont {R.~M.}\ \bibnamefont
  {Lutchyn}}, \bibinfo {author} {\bibfnamefont {J.~D.}\ \bibnamefont {Sau}}, \
  and\ \bibinfo {author} {\bibfnamefont {S.}~\bibnamefont {Das~Sarma}},\
  }\href@noop {} {\bibfield  {journal} {\bibinfo  {journal} {Phys. Rev. Lett.}\
  }\textbf {\bibinfo {volume} {105}},\ \bibinfo {pages} {077001} (\bibinfo
  {year} {2010})}\BibitemShut {NoStop}%
\bibitem [{\citenamefont {Oreg}\ \emph {et~al.}(2010)\citenamefont {Oreg},
  \citenamefont {Refael},\ and\ \citenamefont {von Oppen}}]{Oreg:2010}%
  \BibitemOpen
  \bibfield  {author} {\bibinfo {author} {\bibfnamefont {Y.}~\bibnamefont
  {Oreg}}, \bibinfo {author} {\bibfnamefont {G.}~\bibnamefont {Refael}}, \ and\
  \bibinfo {author} {\bibfnamefont {F.}~\bibnamefont {von Oppen}},\ }\href@noop
  {} {\bibfield  {journal} {\bibinfo  {journal} {Phys. Rev. Lett.}\ }\textbf
  {\bibinfo {volume} {105}},\ \bibinfo {pages} {177002} (\bibinfo {year}
  {2010})}\BibitemShut {NoStop}%
\bibitem [{\citenamefont {Roy}(2010)}]{Roy:2010}%
  \BibitemOpen
  \bibfield  {author} {\bibinfo {author} {\bibfnamefont {R.}~\bibnamefont
  {Roy}},\ }\href@noop {} {\bibfield  {journal} {\bibinfo  {journal} {Phys.
  Rev. Lett.}\ }\textbf {\bibinfo {volume} {105}},\ \bibinfo {pages} {186401}
  (\bibinfo {year} {2010})}\BibitemShut {NoStop}%
\bibitem [{\citenamefont {{Bena}}\ \emph {et~al.}(2011)\citenamefont {{Bena}},
  \citenamefont {{Sticlet}},\ and\ \citenamefont {{Simon}}}]{Bena:2011}%
  \BibitemOpen
  \bibfield  {author} {\bibinfo {author} {\bibfnamefont {C.}~\bibnamefont
  {{Bena}}}, \bibinfo {author} {\bibfnamefont {D.}~\bibnamefont {{Sticlet}}}, \
  and\ \bibinfo {author} {\bibfnamefont {P.}~\bibnamefont {{Simon}}},\
  }\href@noop {} {\bibfield  {journal} {\bibinfo  {journal} {ArXiv e-prints}\ }
  (\bibinfo {year} {2011})},\ \Eprint {http://arxiv.org/abs/1109.5697}
  {arXiv:1109.5697 [cond-mat.mes-hall]} \BibitemShut {NoStop}%
\bibitem [{\citenamefont {DeGottardi}\ \emph {et~al.}(2011)\citenamefont
  {DeGottardi}, \citenamefont {Sen},\ and\ \citenamefont
  {Vishveshwara}}]{DeGottardi:2011}%
  \BibitemOpen
  \bibfield  {author} {\bibinfo {author} {\bibfnamefont {W.}~\bibnamefont
  {DeGottardi}}, \bibinfo {author} {\bibfnamefont {D.}~\bibnamefont {Sen}}, \
  and\ \bibinfo {author} {\bibfnamefont {S.}~\bibnamefont {Vishveshwara}},\
  }\href@noop {} {\bibfield  {journal} {\bibinfo  {journal} {New J. Phys.}\
  }\textbf {\bibinfo {volume} {13}},\ \bibinfo {pages} {065028} (\bibinfo
  {year} {2011})}\BibitemShut {NoStop}%
\bibitem [{\citenamefont {Kitaev}(2001)}]{Kitaev:2001}%
  \BibitemOpen
  \bibfield  {author} {\bibinfo {author} {\bibfnamefont {A.~Y.}\ \bibnamefont
  {Kitaev}},\ }\href@noop {} {\bibfield  {journal} {\bibinfo  {journal}
  {Physics Uspekhi}\ }\textbf {\bibinfo {volume} {44}},\ \bibinfo {pages} {131}
  (\bibinfo {year} {2001})}\BibitemShut {NoStop}%
\bibitem [{\citenamefont {Pfeuty}(1970)}]{Pfeuty:1970}%
  \BibitemOpen
  \bibfield  {author} {\bibinfo {author} {\bibfnamefont {P.}~\bibnamefont
  {Pfeuty}},\ }\href@noop {} {\bibfield  {journal} {\bibinfo  {journal} {Ann.
  Phys. (NY)}\ }\textbf {\bibinfo {volume} {57}},\ \bibinfo {pages} {79}
  (\bibinfo {year} {1970})}\BibitemShut {NoStop}%
\bibitem [{\citenamefont {Read}\ and\ \citenamefont {Green}(2000)}]{Read:2000}%
  \BibitemOpen
  \bibfield  {author} {\bibinfo {author} {\bibfnamefont {N.}~\bibnamefont
  {Read}}\ and\ \bibinfo {author} {\bibfnamefont {D.}~\bibnamefont {Green}},\
  }\href@noop {} {\bibfield  {journal} {\bibinfo  {journal} {Phys. Rev. B}\
  }\textbf {\bibinfo {volume} {61}},\ \bibinfo {pages} {10267} (\bibinfo {year}
  {2000})}\BibitemShut {NoStop}%
\bibitem [{\citenamefont {Alicea}\ \emph {et~al.}(2011)\citenamefont {Alicea},
  \citenamefont {Oreg}, \citenamefont {Refael}, \citenamefont {von Oppen},\
  and\ \citenamefont {Fisher}}]{Alicea:2011}%
  \BibitemOpen
  \bibfield  {author} {\bibinfo {author} {\bibfnamefont {J.}~\bibnamefont
  {Alicea}}, \bibinfo {author} {\bibfnamefont {Y.}~\bibnamefont {Oreg}},
  \bibinfo {author} {\bibfnamefont {G.}~\bibnamefont {Refael}}, \bibinfo
  {author} {\bibfnamefont {F.}~\bibnamefont {von Oppen}}, \ and\ \bibinfo
  {author} {\bibfnamefont {M.}~\bibnamefont {Fisher}},\ }\href@noop {}
  {\bibfield  {journal} {\bibinfo  {journal} {Nature Phys.}\ }\textbf {\bibinfo
  {volume} {7}},\ \bibinfo {pages} {412} (\bibinfo {year} {2011})}\BibitemShut
  {NoStop}%
\bibitem [{\citenamefont {Kopp}\ and\ \citenamefont
  {Chakravarty}(2005)}]{Kopp:2005}%
  \BibitemOpen
  \bibfield  {author} {\bibinfo {author} {\bibfnamefont {A.}~\bibnamefont
  {Kopp}}\ and\ \bibinfo {author} {\bibfnamefont {S.}~\bibnamefont
  {Chakravarty}},\ }\href@noop {} {\bibfield  {journal} {\bibinfo  {journal}
  {Nature Phys.}\ }\textbf {\bibinfo {volume} {1}},\ \bibinfo {pages} {53}
  (\bibinfo {year} {2005})}\BibitemShut {NoStop}%
\bibitem [{\citenamefont {Hirsch}\ and\ \citenamefont
  {Mazenko}(1979)}]{Hirsch:1979}%
  \BibitemOpen
  \bibfield  {author} {\bibinfo {author} {\bibfnamefont {J.~E.}\ \bibnamefont
  {Hirsch}}\ and\ \bibinfo {author} {\bibfnamefont {G.~F.}\ \bibnamefont
  {Mazenko}},\ }\href@noop {} {\bibfield  {journal} {\bibinfo  {journal} {Phys.
  Rev. B}\ }\textbf {\bibinfo {volume} {19}},\ \bibinfo {pages} {2656}
  (\bibinfo {year} {1979})}\BibitemShut {NoStop}%
\bibitem [{\citenamefont {Castro~Neto}\ and\ \citenamefont
  {Fradkin}(1993)}]{Neto:1993}%
  \BibitemOpen
  \bibfield  {author} {\bibinfo {author} {\bibfnamefont {A.~H.}\ \bibnamefont
  {Castro~Neto}}\ and\ \bibinfo {author} {\bibfnamefont {E.}~\bibnamefont
  {Fradkin}},\ }\href@noop {} {\bibfield  {journal} {\bibinfo  {journal} {Nucl.
  Phys. B}\ }\textbf {\bibinfo {volume} {400}},\ \bibinfo {pages} {525}
  (\bibinfo {year} {1993})}\BibitemShut {NoStop}%
\bibitem [{\citenamefont {Anderson}(1958)}]{Anderson:1958}%
  \BibitemOpen
  \bibfield  {author} {\bibinfo {author} {\bibfnamefont {P.~W.}\ \bibnamefont
  {Anderson}},\ }\href@noop {} {\bibfield  {journal} {\bibinfo  {journal}
  {Phys. Rev.}\ }\textbf {\bibinfo {volume} {110}},\ \bibinfo {pages} {827}
  (\bibinfo {year} {1958})}\BibitemShut {NoStop}%
\bibitem [{\citenamefont {Fradkin}\ and\ \citenamefont
  {Susskind}(1978)}]{Fradkin:1978}%
  \BibitemOpen
  \bibfield  {author} {\bibinfo {author} {\bibfnamefont {E.}~\bibnamefont
  {Fradkin}}\ and\ \bibinfo {author} {\bibfnamefont {L.}~\bibnamefont
  {Susskind}},\ }\href@noop {} {\bibfield  {journal} {\bibinfo  {journal}
  {Phys. Rev. D}\ }\textbf {\bibinfo {volume} {17}},\ \bibinfo {pages} {2637}
  (\bibinfo {year} {1978})}\BibitemShut {NoStop}%
\bibitem [{\citenamefont {Barouch}\ and\ \citenamefont
  {McCoy}(1971)}]{Barouch:1971}%
  \BibitemOpen
  \bibfield  {author} {\bibinfo {author} {\bibfnamefont {E.}~\bibnamefont
  {Barouch}}\ and\ \bibinfo {author} {\bibfnamefont {B.~M.}\ \bibnamefont
  {McCoy}},\ }\href@noop {} {\bibfield  {journal} {\bibinfo  {journal} {Phys.
  Rev. A}\ }\textbf {\bibinfo {volume} {3}},\ \bibinfo {pages} {786} (\bibinfo
  {year} {1971})}\BibitemShut {NoStop}%
  \bibitem{Lutchyn:2011} In the following
related publications  multi-critical points in the topological
phase diagram have also been discussed:  R. Lutchyn and M. P. A Fisher,  arXiv:1104.2358 (2011); E. Sela, A. Altland, and A. Rosch, Phys. Rev. B {\bf 84}, 085114 (2011).
\bibitem [{\citenamefont {Wei}\ \emph {et~al.}(2011)\citenamefont {Wei},
  \citenamefont {Vishveshwara},\ and\ \citenamefont {Goldbart}}]{Wei:2011}%
  \BibitemOpen
  \bibfield  {author} {\bibinfo {author} {\bibfnamefont {T.-W.}\ \bibnamefont
  {Wei}}, \bibinfo {author} {\bibfnamefont {S.}~\bibnamefont {Vishveshwara}}, \
  and\ \bibinfo {author} {\bibfnamefont {P.~M.}\ \bibnamefont {Goldbart}},\
  }\href@noop {} {\bibfield  {journal} {\bibinfo  {journal} {Quantum Inf.
  Comput.}\ }\textbf {\bibinfo {volume} {11}},\ \bibinfo {pages} {0326}
  (\bibinfo {year} {2011})}\BibitemShut {NoStop}%
\bibitem [{\citenamefont {Hasan}\ and\ \citenamefont
  {Kane}(2011)}]{Hasan:2011}%
  \BibitemOpen
  \bibfield  {author} {\bibinfo {author} {\bibfnamefont {M.~Z.}\ \bibnamefont
  {Hasan}}\ and\ \bibinfo {author} {\bibfnamefont {C.~L.}\ \bibnamefont
  {Kane}},\ }\href@noop {} {\bibfield  {journal} {\bibinfo  {journal} {Rev.
  Mod. Phys.}\ }\textbf {\bibinfo {volume} {82}},\ \bibinfo {pages} {3045}
  (\bibinfo {year} {2011})}\BibitemShut {NoStop}%
\bibitem [{\citenamefont {{Qi}}\ and\ \citenamefont {{Zhang}}(2010)}]{Qi:2010}%
  \BibitemOpen
  \bibfield  {author} {\bibinfo {author} {\bibfnamefont {X.}~\bibnamefont
  {{Qi}}}\ and\ \bibinfo {author} {\bibfnamefont {S.}~\bibnamefont {{Zhang}}},\
  }\href@noop {} {\bibfield  {journal} {\bibinfo  {journal} {ArXiv e-prints}\ }
  (\bibinfo {year} {2010})},\ \Eprint {http://arxiv.org/abs/1008.2026}
  {arXiv:1008.2026 [cond-mat.mes-hall]} \BibitemShut {NoStop}%
\bibitem [{\citenamefont {Roy}(2011)}]{Roy:2011}%
  \BibitemOpen
  \bibfield  {author} {\bibinfo {author} {\bibfnamefont {R.}~\bibnamefont
  {Roy}},\ }\href@noop {} {} (\bibinfo {year} {2011})\BibitemShut {NoStop}%
\bibitem [{\citenamefont {Ghosh}\ \emph {et~al.}(2010)\citenamefont {Ghosh},
  \citenamefont {Sau}, \citenamefont {Tewari},\ and\ \citenamefont
  {Das~Sarma}}]{Ghosh:2010}%
  \BibitemOpen
  \bibfield  {author} {\bibinfo {author} {\bibfnamefont {P.}~\bibnamefont
  {Ghosh}}, \bibinfo {author} {\bibfnamefont {J.~D.}\ \bibnamefont {Sau}},
  \bibinfo {author} {\bibfnamefont {S.}~\bibnamefont {Tewari}}, \ and\ \bibinfo
  {author} {\bibfnamefont {S.}~\bibnamefont {Das~Sarma}},\ }\href@noop {}
  {\bibfield  {journal} {\bibinfo  {journal} {Physical Review B}\ }\textbf
  {\bibinfo {volume} {82}},\ \bibinfo {pages} {184525} (\bibinfo {year}
  {2010})}\BibitemShut {NoStop}%
\bibitem [{\citenamefont {{Fidkowski}}\ \emph {et~al.}(2011)\citenamefont
  {{Fidkowski}}, \citenamefont {{Lutchyn}}, \citenamefont {{Nayak}},\ and\
  \citenamefont {{Fisher}}}]{Fidkowski:2011}%
  \BibitemOpen
  \bibfield  {author} {\bibinfo {author} {\bibfnamefont {L.}~\bibnamefont
  {{Fidkowski}}}, \bibinfo {author} {\bibfnamefont {R.~M.}\ \bibnamefont
  {{Lutchyn}}}, \bibinfo {author} {\bibfnamefont {C.}~\bibnamefont {{Nayak}}},
  \ and\ \bibinfo {author} {\bibfnamefont {M.~P.~A.}\ \bibnamefont
  {{Fisher}}},\ }\href@noop {} {\bibfield  {journal} {\bibinfo  {journal}
  {ArXiv e-prints}\ } (\bibinfo {year} {2011})},\ \Eprint
  {http://arxiv.org/abs/1106.2598} {arXiv:1106.2598 [cond-mat.str-el]}
  \BibitemShut {NoStop}%
\bibitem [{\citenamefont {Sau}\ \emph {et~al.}(2011)\citenamefont {Sau},
  \citenamefont {Halperin}, \citenamefont {Flensberg},\ and\ \citenamefont
  {Das~Sarma}}]{Sau:2011}%
  \BibitemOpen
  \bibfield  {author} {\bibinfo {author} {\bibfnamefont {J.~D.}\ \bibnamefont
  {Sau}}, \bibinfo {author} {\bibfnamefont {B.~I.}\ \bibnamefont {Halperin}},
  \bibinfo {author} {\bibfnamefont {K.}~\bibnamefont {Flensberg}}, \ and\
  \bibinfo {author} {\bibfnamefont {S.}~\bibnamefont {Das~Sarma}},\ }\href@noop
  {} {\bibfield  {journal} {\bibinfo  {journal} {Physical Review B}\ }\textbf
  {\bibinfo {volume} {84}},\ \bibinfo {pages} {144509} (\bibinfo {year}
  {2011})},\ \bibinfo {note} {pRB}\BibitemShut {NoStop}%
 \bibitem{Fidkowski:2010} L. Fidkowski and A. Kitaev, Phys. Rev. B {\bf 81}, 134509 (2010).
\end{thebibliography}
\end{document}